\documentclass[twocolumn,preprint]{aastex631}

\usepackage{xcolor}

\usepackage{natbib}
\usepackage{graphicx}
\usepackage{amsmath}
\usepackage{booktabs}
\usepackage{xspace}
\usepackage{hyperref}
\usepackage[T1]{fontenc}
\usepackage{lipsum}
\usepackage{comment}
\usepackage{multirow}
\usepackage{enumitem}

\shorttitle{GW from SS-SP Systems}
\shortauthors{Kurban et al.}

\begin{document}

\title{Probing strange quark matter objects with future space-based gravitational wave detectors DECIGO and BBO}

\correspondingauthor{Abdusattar Kurban}
\email{akurban@xao.ac.cn}

\author[0000-0002-2162-0378]{Abdusattar Kurban}
\affil{State Key Laboratory of Radio Astronomy and Technology, Xinjiang Astronomical Observatory, CAS, 150 Science 1-Street, Urumqi, Xinjiang, 830011, People's Republic of China}
\affil{Xinjiang Key Laboratory of Radio Astrophysics, Urumqi 830011, Xinjiang, People's Republic of China}

\author[0000-0003-4686-5977]{Xia Zhou}
\affil{State Key Laboratory of Radio Astronomy and Technology, Xinjiang Astronomical Observatory, CAS, 150 Science 1-Street, Urumqi, Xinjiang, 830011, People's Republic of China}
\affil{Xinjiang Key Laboratory of Radio Astrophysics, Urumqi 830011, Xinjiang, People's Republic of China}

\author[0000-0002-9786-8548]{Na Wang}
\affil{State Key Laboratory of Radio Astronomy and Technology, Xinjiang Astronomical Observatory, CAS, 150 Science 1-Street, Urumqi, Xinjiang, 830011, People's Republic of China}
\affil{Xinjiang Key Laboratory of Radio Astrophysics, Urumqi 830011, Xinjiang, People's Republic of China}

\author[0000-0001-7199-2906]{Yong-Feng Huang}
\affil{School of Astronomy and Space Science, Nanjing University, Nanjing 210023, People's Republic of China}
\affiliation{Key Laboratory of Modern Astronomy and Astrophysics (Nanjing University), Ministry of Education, People's Republic of China}

\author[0000-0002-7662-3875]{Wenming Yan}
\affil{State Key Laboratory of Radio Astronomy and Technology, Xinjiang Astronomical Observatory, CAS, 150 Science 1-Street, Urumqi, Xinjiang, 830011, People's Republic of China}
\affil{Xinjiang Key Laboratory of Radio Astrophysics, Urumqi 830011, Xinjiang, People's Republic of China}

\author[0000-0002-5381-6498]{Jianping Yuan}
\affil{State Key Laboratory of Radio Astronomy and Technology, Xinjiang Astronomical Observatory, CAS, 150 Science 1-Street, Urumqi, Xinjiang, 830011, People's Republic of China}
\affil{Xinjiang Key Laboratory of Radio Astrophysics, Urumqi 830011, Xinjiang, People's Republic of China}

\author[0000-0003-1845-4900]{Ali Esamdin}
\affil{Xinjiang Astronomical Observatory, Chinese Academy of Sciences, Urumqi 830011, Xinjiang, People's Republic of China}

\author{Rai Yuen}
\affil{State Key Laboratory of Radio Astronomy and Technology, Xinjiang Astronomical Observatory, CAS, 150 Science 1-Street, Urumqi, Xinjiang, 830011, People's Republic of China}
\affil{Xinjiang Key Laboratory of Radio Astrophysics, Urumqi 830011, Xinjiang, People's Republic of China}

\author[0000-0002-3525-791X]{Jin-Hong Chen}
\affil{Department of Physics, University of Hong Kong, Pokfulam Road, Hong Kong, People's Republic of China}
\affil{The Hong Kong Institute for Astronomy and Astrophysics, University of Hong Kong, Hong Kong, People's Republic of China}
\affil{Shenzhen Institute of Research and Innovation, The University of Hong Kong, Shenzhen 518057, People's Republic of China}

\author[0009-0000-0467-0050]{Xiao-Fei Dong}
\affil{School of Astronomy and Space Science, Nanjing University, Nanjing 210023, People's Republic of China}

\author[0000-0003-2991-7421]{Zhigang Wen}
\affil{State Key Laboratory of Radio Astronomy and Technology, Xinjiang Astronomical Observatory, CAS, 150 Science 1-Street, Urumqi, Xinjiang, 830011, People's Republic of China}
\affil{Xinjiang Key Laboratory of Radio Astrophysics, Urumqi 830011, Xinjiang, People's Republic of China}

\author[0000-0002-9061-6022]{Yu-Bin Wang}
\affil{School of Physics and Electronic Engineering, Sichuan University of Science \& Engineering, Zigong 643000, People's Republic of China}
\affil{Research Center for Ray Detection Discipline and Technology, Sichuan University of Science \& Engineering, Zigong 643000, People's Republic of China}


\begin{abstract}
The Strange Quark Matter (SQM) hypothesis posits that objects composed of SQM could exist across a wide mass range, from strange planets (SPs) to strange stars (SSs). It has been proposed that gravitational waves (GWs) emitted by inspiraling SS-SP systems may be detectable by ground-based GW observatories such as advanced LIGO and the Einstein Telescope. Nevertheless, such a system may undergo an extended period of orbital evolution in a close configuration before entering the inspiraling phase. During this time, it can generate continuous GW signals at frequencies ranging from milli-hertz (mHz) to deci-hertz (dHz). The detailed characteristics of these GWs have not yet been thoroughly explored. In this study, we delve into the continuous GW features of SS-SP systems, with a focus on exploring the physically viable parameter space. We compared the GW signals emitted by these systems to the sensitivity curves of next-generation space-based GW detectors like the Deci-hertz Interferometer Gravitational wave Observatory (DECIGO) and the Big Bang Observer (BBO). Our analyses demonstrate that both the DECIGO and BBO detectors are capable of detecting continuous GWs from SS-SP systems across a broad parameter space. These GWs carry important information for testing the SQM hypothesis, as well as for advancing our understanding of supernovae and compact star merger processes.

\end{abstract}


\section{Introduction} \label{sec:intro}

The Bodmer-Witten hypothesis posits strange quark matter (SQM) as the ground state of hadronic interactions \citep{Bodmer1971,Witten1984PhRvD}. It can exist in the form of astrophysical objects with masses ranging from planetary mass to that of pulsars (PSRs). Pulsars might be conventional neutron stars (NSs), or they could be strange quark stars (strange stars; SSs) made up of SQM consists of an equal number of up, down, and strange quarks \citep{Farhi1984PhRvD,Witten1984PhRvD}. Furthermore, strange stars may exist in either a bare state or be covered by a crust of normal matter \citep{Haensel1986AA,Alcock1986ApJ,2022PhLB..83237204K,2024FrASS..1109463Z}. It was argued that strange stars could coexist with NSs \citep{2009PhRvL.103a1101B,2018ApJ...852L..32D}. Nevertheless, the observational differentiation between these two classes of compact stars remains a subject area of ongoing research. 

It has long been suggested that a strange star can form during catastrophic events. On the one hand, strange stars may form through the hadron-quark phase transition in the supernova process or shortly after, when the proto-neutron star reaches sufficiently high densities \citep{1995ApJ...440..815D,2009PhRvL.102h1101S}. Further numerical simulations validate this formation mechanism \citep{2011ApJS..194...39F,2013A&A...558A..50N,2020PhRvL.125e1102Z}. The sources RX J1856.5$-$3754, HESS J1731$-$347, and PSR J0205+6449 may be formed in this way. Analyses of the data acquired by the X-ray and optical observations of RX J1856.5$-$3754 imply the existence of exotic matter inside, suggesting it could be a strange star \citep{Drake_2002,Xu_2002}. It was argued that the central compact object in the supernova remnant HESS J1731$-$347 \citep{2022NatAs...6.1444D} might be a strange star \citep{2024ApJ...967..159D,2025RAA....25e5016Y}, or a hybrid star \citep{2026PhRvD.113d4002C}. PSR J0205+6449 situated in the supernova remnant 3C58 is regarded as a potential source characterized by an exotic core \citep{2024NatAs...8.1020M}. It has also been proposed that an existing massive NS can undergo a phase transition to convert into a strange star \citep{1996PhRvL..77.1210C,2025arXiv251008707Z}. On the other hand, a strange star may form via a hadron-quark phase transition in a binary NS merger. Results from numerical relativity simulations \citep{2020PhRvL.124q1103W,2020PhRvL.125n1103B,2026PhRvD.113d4057H} show that the hadron-quark phase transition may take place in the process of binary NS merger, ultimately resulting in the formation of a strange star.

Recent mass measurements of several high-mass pulsars PSR J1614$-$2230 \cite[$1.97\pm0.04 \, {\rm M_{\odot}}$;][]{2010Natur.467.1081D}, PSR J0348+0432 \cite[$2.01 \pm 0.04 \, {\rm M_{\odot}}$;][]{2013Sci...340..448A}, PSR J0740+6620 \cite[$2.08\pm0.07 \, {\rm M_{\odot}}$;][]{2021ApJ...918L..28M}, and PSR J0952$-$0607 \cite[$2.35 \pm 0.17 \, {\rm M_{\odot}}$;][]{2022ApJ...934L..17R} provide an opportunity to study the internal composition of these enigmatic compact stars. The existence of quark-matter cores in such massive pulsars is proposed with some evidence \citep{2020NatPh..16..907A,2023NatCo..14.8451A,2026A&A...706A.203S}. A study based on the X-ray timing and spectral analysis utilizing NICER data indicated that PSR J0614$-$3329 is a potential candidate for a strange star \citep{2025arXiv250802652S}. In particular, the gravitational wave (GW) event GW190814 \citep{2020ApJ...896L..44A} shows the existence of an interesting compact object with a mass of $2.6 \, {\rm M_{\odot}}$. Some authors argued that this massive object could be a strange star \citep{2021PhRvL.126p2702B} or a rapidly rotating NS with an exotic SQM core \citep{2021PhRvC.103b5808D} or a supermassive and superfast pulsar \citep{2021ApJ...910...62Z}, or likely to be a light black hole \citep{2020PhRvC.102f5805F}. The total mass of the components associated with GW190425 \citep{2020ApJ...892L...3A} is reported to be $3.4^{+0.3}_{-0.1} \, {\rm M_{\odot}}$. Calculations by \cite{2022PhLB..83337388S} suggest that this remnant of compact binary coalescence may be a strange star. At present, the exact internal composition of these compact objects still remains an enigma that is extensively explored.

Previous studies have made great efforts to distinguish strange stars from NSs by analyzing the differences in their mass-radius relations, maximum masses, rotation rates, and cooling properties (see the analyses and relevant references contained in \cite{Kuerban2020ApJ}). When concerning the measurement errors in observational data and uncertainties in model parameters, the differences between strange stars and NSs are not significant, so distinguishing between these two types of compact stars with different internal structures faces difficulties. It has further been suggested that strange stars and NSs can be distinguished by comparing their differing GW characteristics in the high-frequency band. Such characteristics are present in rotating strange stars \citep{2002MNRAS.337.1224A} and rotating NSs \citep{2025FrASS..1225459L,2026ApJ...999..262L}, binary strange star mergers and binary NS mergers \citep{2010PhRvD..81b4012B}, and f-mode oscillation frequencies of NSs with quark matter \citep{2026JCAP...03..017S}. The existence of strange stars formed from a hadron-quark phase transition in binary NS mergers can be diagnosed through detection of post-merger signatures in GW frequencies \citep{2019PhRvL.122f1102B,2020PhRvL.124q1103W,2020PhRvL.125n1103B,2026PhRvD.113d4057H}. These GWs are expected to be detected by more advanced ground-based GW observatories in the future. While the methodologies discussed above have substantially improved our comprehension of this field, the confirmation of strange stars still requires future multi-messenger investigations.

Encouragingly, a unique celestial system composed of a strange star and a strange planet (i.e., an SS-SP system) is expected to be a powerful new probe for exploring the equation of state describing the internal composition and structure of pulsars. Strong GWs at high-frequency can be emitted in their final inspiral phase. Specifically, these events occur when the orbital separation between the two compact objects decreases to less than twenty times the tidal disruption radius, and the resulting GW bursts are detectable by ground-based GW telescopes such as advanced LIGO and the Einstein Telescope \citep{Geng2015ApJ_a,Zhang2024MNRAS}. The direct detection of GWs from a binary black hole merger \citep{2016PhRvL.116f1102A} and a binary NS merger \citep{Abbott2017PhRvL.119p1101A} sheds light on the possibility for detecting an SS-SP system through GW astronomy. 

However, before the strange planet enters the spiraling-in phase, it could undergo a long-term orbital evolution in a close orbit \citep{Huang2017ApJ,Kuerban2019AIPC,Kuerban2020ApJ}. During this process, continuous GWs with frequencies lower than those of the spiraling-in process are expected. The frequency range of such GWs that is inaccessible to ground-based detectors can be covered by the frequency bands of the space-based GW detectors designed to observe low-frequency GWs. Space-based GW detectors, including the Laser Interferometer Space Antenna \citep[LISA;][]{Amaro-Seoane2017arXiv170200786A}, TianQin \citep{Luo2016CQGra}, and Taiji \citep{Ruan2018arXiv180709495R}, aim to detect GWs at $\text{mHz}$ frequency range \citep{Ruan2020IJMPA,Huang2020PhRvD,Amaro-Seoane2023LRR}, thus tracing the history of compact object mergers. Our previous analysis indicated that GWs in this frequency band, produced by a system consisting of a compact star and a planetary-mass companion with a mass less than one Jupiter mass (\({\rm M_{Jup}} = 9.55\times10^{-4} \, {\rm M_{\odot}}\)), are unlikely to be detected by these near-term detectors \citep{Kurban2026AA}. Notably, two proposed projects, the Deci-hertz Interferometer Gravitational wave Observatory \citep[DECIGO;][]{Kawamura2006CQGra} and the Big Bang Observer \citep[BBO;][]{Harry_2006}, have been designed to detect deci-hertz GWs originating from the early Universe and compact objects \citep{Cutler2006PhRvD,Yagi2011PhRvD}, aiming to fill the frequency gap between LISA and ground-based GW detectors. They achieve higher detection sensitivity than LISA, TianQin, and Taiji within this frequency band. These properties of DECIGO and BBO enable a novel approach to testing the SQM hypothesis through the observation of low-frequency continuous GWs produced by SS-SP systems. In this study, we thoroughly investigate the characteristics and observability of such low-frequency continuous GWs from SS-SP systems, in the hope of providing new clues and methods for a deeper understanding of the internal structure of pulsars. 

The structure of this paper is organized as follows. The theoretical framework for GWs is given in Section \ref{sec:signal model}. The parameter sets for SS-SP systems are presented in Section \ref{sec:para_set}. In Section \ref{sec:results}, we report the numerical outcomes obtained in this study, which characterize the GW properties and the observable parameter space of SS-SP systems. Finally, Section \ref{sec:conclusion_discussion} presents our conclusions and some brief discussion.


\section{GW signal model}\label{sec:signal model}

For a binary system composed of a primary strange star mass $m_1$, a strange planet mass $m_2$, and semimajor axis $a$, its orbital frequency $f_{\rm orb} $ can be expressed as
\begin{equation} \label{eq:f_orb}
	f_{\rm orb} = \frac{1}{2\pi} \sqrt{\frac{G(m_1 + m_2)}{a^3}}.
\end{equation}
The energy loss rate from the system due to the GW radiation is \citep{Peters1963}
\begin{equation} \label{eq:edot}
	\dot{E} = \sum_{n = 1}^{\infty} \dot{E}_n = \sum_{n = 1}^{\infty} \frac{32}{5} \frac{G^{7 / 3}}{c^{5}}\left(2 \pi f_{\mathrm{orb}} \mathcal{M}_{c}\right)^{10 / 3} g(n, e),
\end{equation}
where $\dot{E}_n$ is the GW power radiated in the $n^{\rm th}$ harmonic over one period of the elliptical motion with eccentricity $e$, $G$ is the gravitational constant, $c$ is the speed of light, $\mathcal{M}_c \equiv (m_1 m_2)^{3/5}/(m_1 + m_2)^{1/5}$ is the chirp mass. 
The function $g(n, e)$ is defined as
\begin{equation} \label{eq:g(n,e)}
	\begin{aligned}
		g(n, e) = \frac{n^{4}}{32} & \Big\{ \Big[ J_{n-2}(n e)-2 e J_{n-1}(n e)+\frac{2}{n} J_{n}(n e) \\
		&\left. +2 e J_{n+1}(n e)-J_{n+2}(n e)\Big]^{2}\right.\\
		&\left.\kern-\nulldelimiterspace\right.
		\!\!\begin{aligned}
			+\big(1-e^{2}\big)\Big[&J_{n-2}(n e)-2 J_{n}(n e) \\
			&+J_{n+2}(n e)\Big]^{2}
		\end{aligned}\\
		&+\frac{4}{3 n^{2}}\Big[J_{n}(n e)\Big]^{2}\Big\},
	\end{aligned}
\end{equation}
where $J_{n}(x)$ is the Bessel function of the first kind. Thus, the \textit{sum} of $g(n, e)$ over all harmonics gives the factor by which the GW emission is enhanced for a binary of eccentricity $e$ over an equivalent circular binary. This enhancement factor is \citep{Peters1963}
\begin{equation} \label{eq:eccentricity_enhancement_factor}
	F(e) = \sum_{n = 1}^{\infty} g(n, e) = \frac{1 + \frac{73}{24} e^2 + \frac{37}{96} e^4}{(1 - e^2)^{7/2}}.
\end{equation}

For a system in an eccentric orbit, the frequency range of GW is wide and can be determined by the harmonics $n$ and orbital frequency $f_{\rm orb}$. The $n^{\rm th}$ harmonic frequency of GW emission is
\begin{equation} \label{eq:nth_freq}
	f_{\text{gw},n} = n f_{\rm orb}.
\end{equation}
For a GW source at a luminosity distance $D_L$, the characteristic strain amplitude of GWs in the $n^{\rm th}$ harmonic is \citep{Wagg2022ApJS}
\begin{equation}\label{eq:char_strain}
	h_{c,n}^2 = \frac{2^{5/3}}{3 \pi^{4/3}} \frac{(G \mathcal{M}_c)^{5/3}}{c^3 D_L^2} \frac{1}{f_{\rm orb}^{1/3}} \frac{g(n, e)}{n F(e)}. 
\end{equation}
For a detector with a sensitivity curve of $S_{\rm n}$, the signal-to-noise ratio (\(\text{S/N}\)) of the GW produced by a system in an eccentric orbit can be estimated by summing the values of \(\text{S/N}\) of all harmonics, which is expressed as
\begin{equation}\label{eq:snr_general}
	\langle \text{S/N} \rangle^2 = \sum_{n = 1}^{\infty} \langle \text{S/N}_n \rangle^2 = \sum_{n = 1}^{\infty} \int_{f_{\text{gw},n}^{i}}^{f_{\text{gw},n}^{f}} d f_{\text{gw},n} \frac{h_{c, n}^2}{f_{\text{gw},n}^2 S_{\rm n}(f_{\text{gw},n})},
\end{equation}
where $f_{\text{gw},n}^{i}$ and $f_{\text{gw},n}^{f}$ represent the initial and final GW frequencies in the $n^{\rm th}$ harmonic. For a given evolution time, these frequencies can be obtained by coupling Equation (\ref{eq:nth_freq}) with the initial and final orbital frequencies calculated from the evolution equations of the orbital frequency and eccentricity. The evolution of these orbital elements under the GW radiation proceeds as follows,
\begin{equation} \label{eq:f_orb_evo}
	\frac{df_{\rm orb}}{dt} = \frac{96}{5\pi} \frac{(G \mathcal{M}_c)^{5/3}}{c^5} (\pi f_{\rm orb})^{11/3} F(e),
\end{equation}
\begin{equation} \label{eq:ecc_evo}
	\frac{de}{dt} = -\frac{304}{15} \frac{(G \mathcal{M}_c)^{5/3}}{c^5} (\pi f_{\rm orb})^{8/3} H(e),
\end{equation}
where
\begin{equation} \label{eq:constant}
	H(e) = \frac{e}{(1 - e^2)^{5/2}} \left(1 + \frac{121}{304} e^2\right).
\end{equation} 

The amplitude spectral density ($\text{ASD}$) represents the amplitude of the GW signal at different frequencies. For $n^{\rm th}$ harmonic, it is $\text{ASD}_n$ and can be expressed as
\begin{equation} \label{eq:asd}
	\text{ASD}_n = \langle \text{S/N}_n \rangle \sqrt{S_{\rm n}(f_{\text{gw},n})}.
\end{equation}

For systems in circular orbits, $g(n, 0) = 1$ for $n = 2$ and $g(n, 0) = 0$ for $n \neq 2$, which means that $F(0) = 1$ \citep{Hamers2021RNAAS}. By substituting these values into Equation (\ref{eq:char_strain}) and Equation (\ref{eq:snr_general}), one can obtain $\text{S/N}$ for an evolving circular system.


\section{Sets of system parameters} \label{sec:para_set}

In this study, we assume that pulsars are strange stars. The mass range of these strange stars, as constrained by electromagnetic observations, lies between approximately $1.4 \, {\rm M_{\odot}}$ and $2.35 \, {\rm M_{\odot}}$ \citep{2022ApJ...934L..17R}. The GW observations have constrained the mass of the compact object to up to approximately $3.4 \, {\rm M_{\odot}}$ \citep{2020ApJ...892L...3A}. Nevertheless, the nature of such objects remains a subject of debate. Consequently, the two typical mass values for host strange stars, $m_1 = \{1.4, 2.0\} \, {\rm M_{\odot}}$, are considered for the calculations in this study.

For distances, we set the parameter space based on the observed distances of pulsars. According to ATNF Pulsar Catalogue\footnote{\url{https://www.atnf.csiro.au/research/pulsar/psrcat/}} \citep{Manchester2005AJ}, the distances of pulsars range from $D_L = 0.092 \, \text{kpc}$ (PSR J1015$-$5359) to $D_L = 100 \, \text{kpc}$ (PSR J0518$-$6939 and PSR J0048$-$7317). Considering a broader range of possible detectable parameter space, we extend the distance to $D_L = 1000 \, \text{kpc}$. This distance covers the distances of most Local Group Galaxies, including the Andromeda Galaxy (M31), which has a distance of $\sim 770 \, \text{kpc}$ ($2.5\times10^6$ light-years\footnote{\url{https://science.nasa.gov/mission/hubble/science/explore-the-night-sky/hubble-messier-catalog/messier-31/}}). Therefore, we set the distance of SS-SP systems as $D_L \in [0.1,1000] \, \text{kpc}$.

We established a potential parameter space for the SS-SP system based on the tidal disruption radius. The primary characteristics of strange planets, when compared with those composed of normal matter, is their capacity to withstand the tidal force exerted by the host strange star at short distances. The disruption radius of an object with a mean density of $\bar{\rho}$ around a compact object mass of $M$ is given as  \cite{Hills1975Natur}
\begin{equation} \label{eq:rtd}
	r_{\text{td}} \approx \left( \frac{6M}{\pi\bar{\rho}}\right)^{\frac{1}{3}}.
\end{equation}
According to this formula, the tidal disruption distances of a normal-matter planet and a strange planet differ significantly owing to their internal compositions. For a normal-matter planet with a mean density of \(\bar{\rho} = 30\) g $\text{cm}^{-3}$, the tidal disruption radius is \(r_{\text{td}} \approx 5.6 \times10^{10} \, \text{cm}\) \citep{Huang2017ApJ,Kuerban2019AIPC,2023MNRAS.522.4265K}. Tidal disruption of low-mass objects (\(\bar{\rho} < 30\) g cm$^{-3}$) around pulsar-like compact stars may produce fast radio bursts \citep{Geng2021Innov,2022ApJ...928...94K,2024EPJC...84..210N} or X-ray bursts \citep{2024A&A...686A..87K}. In contrast, for a strange planet with a mean density of \(\bar{\rho} \approx 4 \times 10^{14}\) g $\text{cm}^{-3}$, the tidal disruption radius is \(r_{\text{td}} \approx 2.37 \times10^{6} \, \text{cm}\) \citep{Kuerban2020ApJ}. As mentioned earlier, the observable events for such close-in systems could be GWs.

Based on the mass-radius relation of SQM objects \citep{Kettner1995PhRvD}, the mass of strange planets can span from that of asteroids (or even smaller bodies) to the mass scale of planetary-mass objects. For a planetary-mass object with an orbital separation of approximately \(5.6\times10^{10} \, \text{cm}\) and a mass exceeding one Jupiter mass (\({\rm M_{Jup}} = 9.55\times10^{-4}\approx10^{-3}\, {\rm M_{\odot}}\)), an alternative interpretation involving a light white dwarfs also exists \citep{Kurban2026AA}. Therefore, the upper bound of the mass range for strange planets considered in this work is set at approximately \(10^{-3}\, {\rm M_{\odot}}\), i.e., \(m_2 \leq 10^{-3}\, {\rm M_{\odot}}\).

\begin{figure}[tbh!]
	\centering
	\plotone{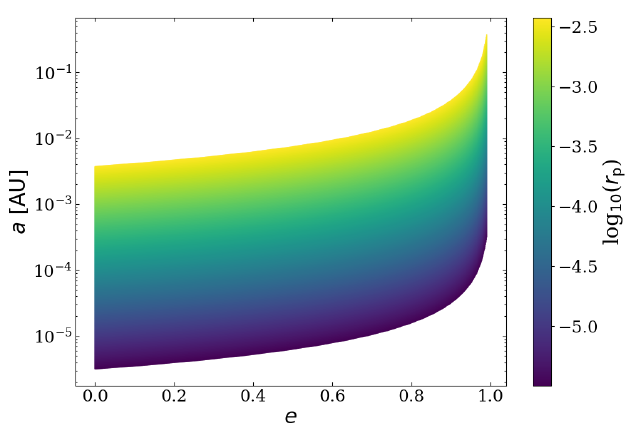}
	\caption{Parameter space for the orbital elements of SS-SP systems considered in this work.}
	\label{fig:fig1}
\end{figure}

The GW properties of SS-SP systems with orbital radius smaller than $20r_{\rm td} \, (\approx 4.75 \times10^{7}\, \text{cm})$ has been investigated in the cases of ground-based GW detectors \citep[e.g.,][]{Geng2015ApJ_a,Zhang2024MNRAS}. Through the combined analysis of this distance together with the aforementioned tidal disruption distance estimates, we are able to determine the range of orbital separation between the strange star and the strange planet. For a system in a circular orbit, their separation can be \(r \in [4.75 \times10^{7}, 5.6 \times10^{10}] \, \text{cm}\). For an eccentric orbit, it is equivalent to the periastron distance \(r_{\text{p}} \in [4.75 \times10^{7}, 5.6 \times10^{10}]\, \text{cm}\). One can constrain the parameter space for the eccentricities and semimajor axes of SS-SP systems using the relation \(r_{\text{p}} = a (1 - e)\).

A recent analysis by \cite{2026ApJ..1000L...2M} inferred that the binary system involving GW200105 has an eccentricity of approximately 0.145, an orbital period of 0.1 seconds, and an unequal mass ratio. This result demonstrates that the GW detection of SS-SP systems characterized by eccentric, ultra-short-period orbits is also possible if such systems exist. Strange planets in circular or eccentric orbits have the potential to form through the following processes. First, an SS-SP system in an eccentric orbit is expected to form under the combined influence of the velocities of both the ejected SQM clumps and the host strange star, which can be formed during explosive events (see Section \ref{sec:conclusion_discussion} for more details). Second, survived strange planets that formed at early Universe \citep{Witten1984PhRvD,1994PhRvL..73.1328C,2026arXiv260316686Q} may have been captured by compact objects, leading to the formation of eccentric orbits with $e\sim1$. Therefore, it is reasonable to take $e \in [0, 0.99]$ for SS-SP systems. Figure \ref{fig:fig1} shows the possible range of the parameter space for SS-SP systems explored in this study.

\begin{table}[tbh!]
	\tabcolsep=4mm
	\small
	\caption{The value range of the parameters of SS-SP systems considered in this study.}
	\label{tab:table_1}
	\begin{tabular}{ll}
		\hline\hline
		Parameter & Value range  \\
		\hline
		\noalign{\smallskip}
		$m_1$ & $\{1.4, 2\}$ ${\rm M_{\odot}}$  \\
		$m_2$ & [$10^{-10}, 10^{-3}$] ${\rm M_{\odot}}$ \\
		$\mathcal{M}_c$ & [$1.144\times10^{-6}, 2.1\times10^{-2}$] ${\rm M_{\odot}}$ \\
		$r_{\rm p}$ & [$4.75\times10^{7}, 5.6\times10^{10}$] $\text{cm}$ \\
		$e$ & $\{0, 0.5, 0.95\}$ \\
		$D_{L}$  & $\{0.1, 10, 1000\}$ $\text{kpc}$ \\	
		\hline
		\noalign{\smallskip}
	\end{tabular}
	\begin{minipage}{0.98\linewidth}
		{Note: The range of $\mathcal{M}_c$ is obtained using the upper and lower bounds of $m_1$ and $m_2$.}	
	\end{minipage}
\end{table}

\begin{figure}[tbh!]
	\centering
	\plotone{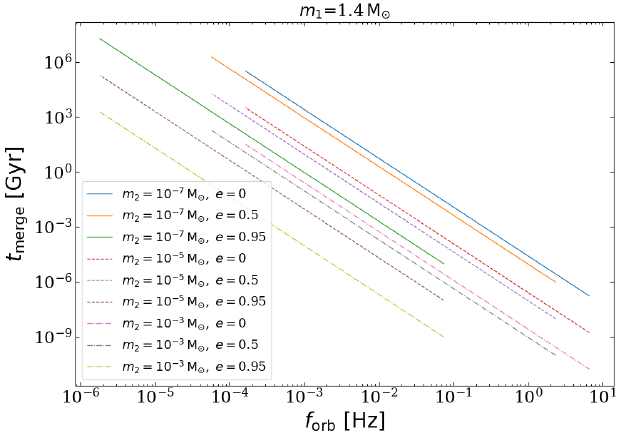}
	\caption{The merger time as a function of orbital frequency, eccentricity, and planet mass. For each curve, the upper and lower limits of the orbital frequency correspond to the periastron distances of \(r_{\rm p} = 4.75 \times10^{7} \, \text{cm}\) and \(r_{\rm p} = 5.6 \times10^{10} \, \text{cm}\), respectively.}
	\label{fig:fig2}
\end{figure}

Figure \ref{fig:fig2} depicts the merger time as a function of orbital frequency, orbital eccentricity, and planet mass. The orbital frequency is derived by integrating the orbital eccentricity, planet mass, and periastron distance. For each line, the upper and lower limits of the orbital frequency correspond to the periastron distances \(r_{\rm p} = 4.75 \times10^{7} \, \text{cm}\) and \(r_{\rm p} = 5.6 \times10^{10} \, \text{cm}\), respectively. As can be discerned from the figure, for a given \( m_1 \) and \(r_{\rm p}\), the orbital frequency and merger time tend to decrease and increase, respectively, with the increase in eccentricity. Conversely, for a given \( e \) and \(r_{\rm p}\), the merger time decreases as the planet mass increases. In most of the configurations, the merger time is sufficiently large when compared with the mission time, indicating that it is feasible to observe the SS-SP system within a wide parameter space. The following section will explore the GW properties of exemplary SS-SP systems and examine the parameter space within which SS-SP systems can be detected by the DECIGO and BBO.

\section{Results}\label{sec:results}

As mentioned in Section \ref{sec:para_set}, planetary systems with orbits that fall within the parameter space illustrated in figure \ref{fig:fig1} can be considered as SS-SP systems. 
To understand the GW characteristics of such systems and constrain the detectable parameter space, we estimate the GW frequency \(f_{\rm gw}\), signal-to-noise ratio $\text{S/N}$, and amplitude spectral density ASD using the parameter sets provided in table \ref{tab:table_1}.

\begin{figure}[tbh!]
	\centering
	\plotone{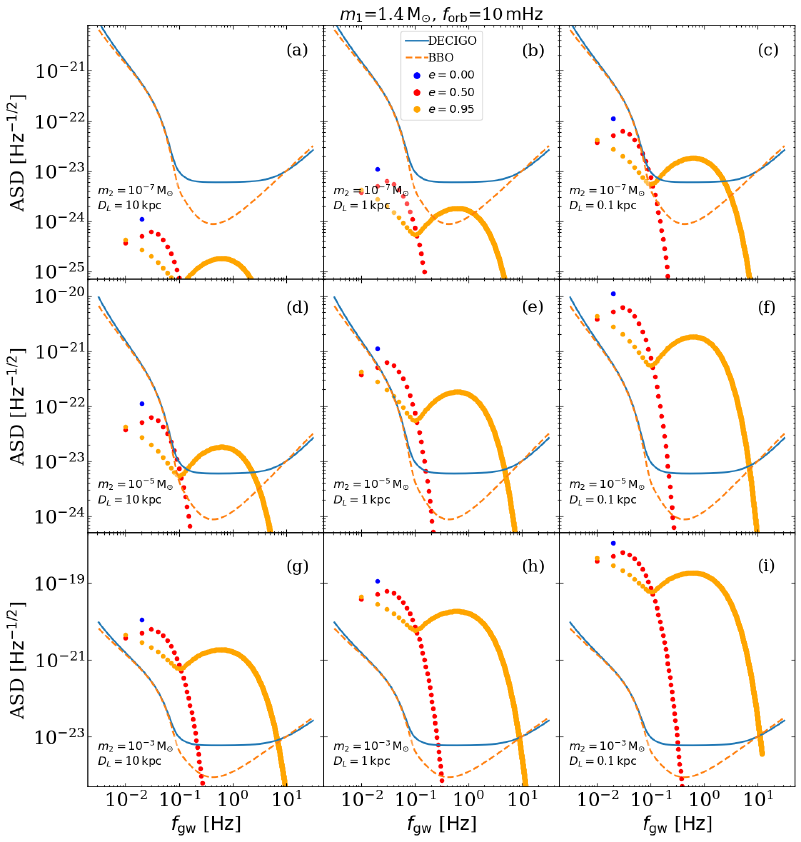}
	\caption{Comparison of the signal calculated for $ T_{\rm obs} = 4 \, \text{years}$ with $\sqrt{S_{\rm n}(f_{\rm gw})}$ at the amplitude spectral density ASD versus the GW frequency $f_{\rm gw}$ plane, where $ S_{\rm n}(f_{\rm gw}) $ represents the noise power spectral densities \citep{Yagi2011PhRvD,Yagi2017PhRvD,Sun2024AA} for DECIGO (solid line) and BBO (dashed line). The subplots represent the calculation results for systems with different combinations across the strange planet mass $m_2 = \{10^{-7}, 10^{-5}, 10^{-3}\} \, {\rm M_{\odot}}$ and luminosity distance $D_L = \{0.1, 1, 10\} \, \text{kpc}$, with the fixed strange star mass $ m_1 = 1.4 \, {\rm M_{\odot}} $ and eccentricity values $ e = 0 $ (blue), $ e = 0.5 $ (red), and $ e = 0.95 $ (orange).}
	\label{fig:ads_ecc_339}
\end{figure}

\begin{table*}
	\tabcolsep=2.0mm
	\small
	\caption{The $\text{S/N}$ values of the exemplary cases considered in this study, with $ T_{\rm obs} = 4 \, \text{years}$.}
	\label{tab:table_2}
	\begin{tabular}{lccccccccc}
		\hline\hline
		$f_{\rm orb}$ & $m_1$ & $m_2$ & $D_L$ & $\text{S/N}_{e=0}$ & $\text{S/N}_{e=0}$ & $\text{S/N}_{e=0.5}$ & $\text{S/N}_{e=0.5}$ & $\text{S/N}_{e=0.95}$ & $\text{S/N}_{e=0.95}$ \\
		$(\text{Hz})$ & $({\rm M_{\odot}})$ & $({\rm M_{\odot}})$ & $(\text{kpc})$  & DECIGO & BBO & DECIGO & BBO & DECIGO & BBO \\
		\hline
		\noalign{\smallskip}
		0.01 & 1.4 & 1.0e-07 & 10 & 0.00 & 0.00 & 0.02 & 0.04 & 0.34 & 1.76 \\
		0.01 & 1.4 & 1.0e-07 & 1 & 0.02 & 0.02 & 0.22 & 0.39 & 3.42 & 17.62 \\
		0.01 & 1.4 & 1.0e-07 & 0.1 & 0.20 & 0.20 & 2.17 & 3.88 & 34.19 & 176.23 \\
		0.01 & 1.4 & 1.0e-05 & 10 & 0.20 & 0.20 & 2.17 & 3.88 & 34.19 & 176.24 \\
		0.01 & 1.4 & 1.0e-05 & 1 & 1.96 & 2.05 & 21.74 & 38.83 & 341.93 & 1762.42 \\
		0.01 & 1.4 & 1.0e-05 & 0.1 & 19.64 & 20.49 & 217.37 & 388.31 & 3419.25 & 17624.16 \\
		0.01 & 1.4 & 1.0e-03 & 10 & 19.64 & 20.49 & 217.32 & 388.23 & 3450.69 & 17783.27 \\
		0.01 & 1.4 & 1.0e-03 & 1 & 196.39 & 204.87 & 2173.24 & 3882.26 & 34506.90 & 177832.70 \\
		0.01 & 1.4 & 1.0e-03 & 0.1 & 1963.88 & 2048.70 & 21732.40 & 38822.56 & 345068.97 & 1778326.99 \\
		\hline
		0.01 & 2.0 & 1.0e-07 & 10 & 0.00 & 0.00 & 0.03 & 0.05 & 0.43 & 2.24 \\
		0.01 & 2.0 & 1.0e-07 & 1 & 0.02 & 0.03 & 0.28 & 0.49 & 4.34 & 22.35 \\
		0.01 & 2.0 & 1.0e-07 & 0.1 & 0.25 & 0.26 & 2.76 & 4.93 & 43.37 & 223.53 \\
		0.01 & 2.0 & 1.0e-05 & 10 & 0.25 & 0.26 & 2.76 & 4.93 & 43.37 & 223.56 \\
		0.01 & 2.0 & 1.0e-05 & 1 & 2.49 & 2.60 & 27.57 & 49.26 & 433.72 & 2235.57 \\
		0.01 & 2.0 & 1.0e-05 & 0.1 & 24.92 & 25.99 & 275.72 & 492.55 & 4337.21 & 22355.65 \\
		0.01 & 2.0 & 1.0e-03 & 10 & 24.91 & 25.99 & 275.68 & 492.47 & 4388.56 & 22615.58 \\
		0.01 & 2.0 & 1.0e-03 & 1 & 249.12 & 259.88 & 2756.81 & 4924.75 & 43885.58 & 226155.76 \\
		0.01 & 2.0 & 1.0e-03 & 0.1 & 2491.23 & 2598.83 & 27568.14 & 49247.48 & 438855.81 & 2261557.55 \\
		\noalign{\smallskip}		
		\hline
	\end{tabular}
\end{table*}

\subsection{Amplitude spectral density and Signal-to-noise ratio}\label{subsec:asd and snr}

We perform S/N and ADS calculations on a set of exemplary cases. For these cases, the parameter settings are as follows: $m_1 = 1.4 \, {\rm M_{\odot}}$, $m_2 = \{10^{-7}, 10^{-5}, 10^{-3}\} \, {\rm M_{\odot}}$, $D_L = \{0.1, 10, 1000\} \, \text{kpc}$, and $e = \{0, 0.5, 0.95\}$. The \(\text{S/N}\) values of SS-SP systems with varying configurations, grouped by the aforementioned parameters, are obtained using the \emph{snr.py} module in LEGWORK \citep{Wagg2022ApJS}, with the effects of orbital evolution fully incorporated in the computation. Within this code file \emph{snr.py}, the evolution time is formulated as $t_{\rm evo} = min(T_{\rm obs}, t_{\rm merge} - t_{\rm before})$. In this expression, $T_{\rm obs}$ corresponds to the observation duration, $t_{\rm merge}$ refers to the merger time, and $t_{\rm before}$ indicates the pre-merger time interval. For systems in circular orbits, $t_{\rm before}$ is set to $1 \, \text{second}$, while for systems in eccentric orbits, $t_{\rm before}$ is assigned a value of $0.1 \, \text{years}$. For our calculations, the power spectral density (PSD) for LISA in the code file \emph{snr.py} is substituted with the corresponding PSDs for DECIGO and BBO, which are obtained from \cite{Yagi2011PhRvD,Yagi2017PhRvD}. Confusion noise is incorporated into these PSDs in accordance with the approach outlined in \cite{Sun2024AA}. For eccentric orbits, the \(\text{S/N}\) values are computed up to \(n = 1200\) harmonics.

\begin{figure*}[th!]
	\centering
	\plotone{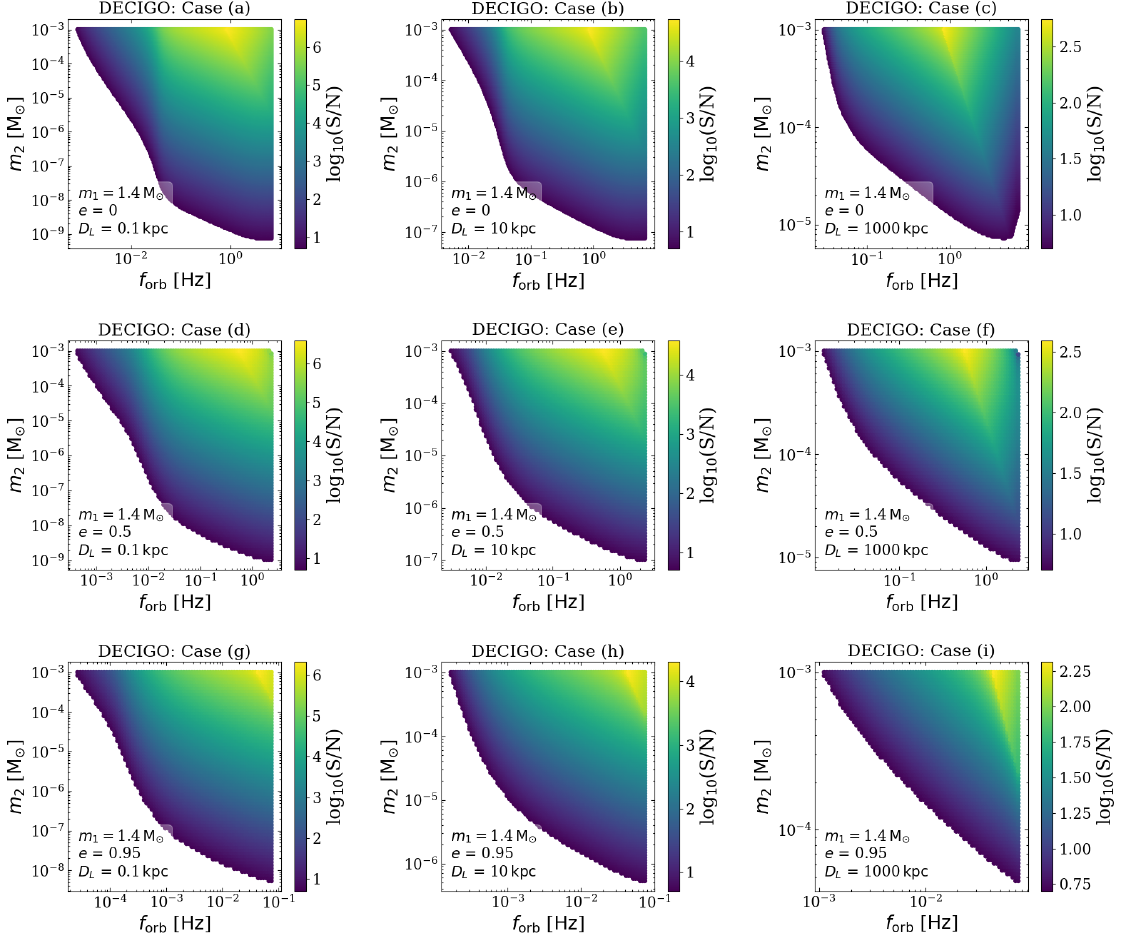}
	\caption{The parameter space in which the SS-SP systems can produce detectable signals with $\text{S/N} \geq 5$, assuming $T_{\rm obs} = 4$ years of observations using DECIGO. Each subplot shows the systems, with a fixed primary mass $m_1$, orbital eccentricity $e$, and distance $D_L$, in the strange planet mass $m_2$ versus the orbital frequency $f_{\rm orb}$ plane. From top to bottom, the eccentricity is taken as $e = 0$, $e = 0.5$, and $e = 0.95$. From left to right, the distance is taken as $D_L = 0.1 \, \text{kpc}$, $D_L = 10 \, \text{kpc}$, and $D_L = 1000 \, \text{kpc}$. The upper bounds of the orbital frequency for each subplot correspond to \(r_{\text{p}} = 4.75 \times10^{7}\, \text{cm}\).}
	\label{fig:fig_para_snr_decigo}
\end{figure*}

\begin{figure*}[th!]
	\centering
	\plotone{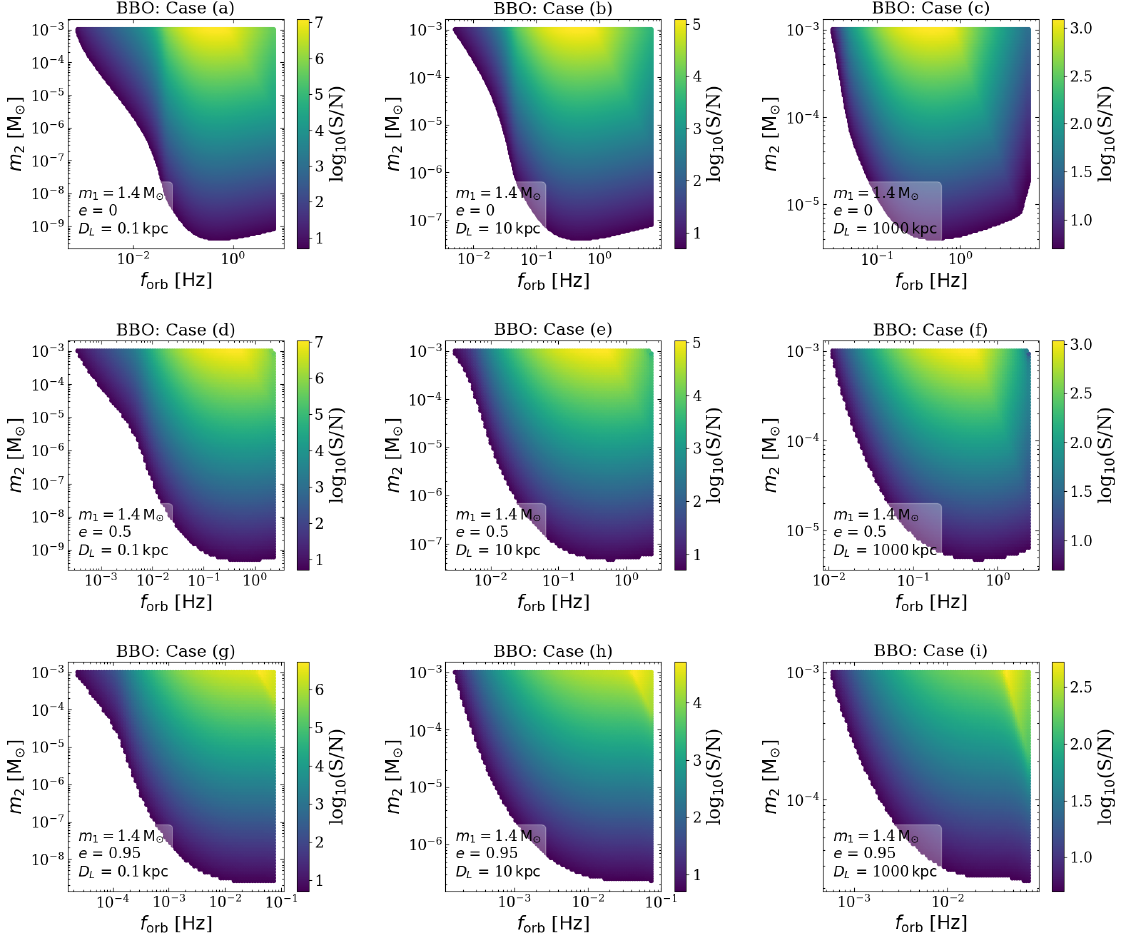}
	\caption{Analogous to the scenario presented in figure \ref{fig:fig_para_snr_decigo}, but for BBO.}
	\label{fig:fig_para_snr_bbo}
\end{figure*}

We compare the \(\text{ASD}\) calculated for various groups of parameters with the sensitivity curves of DECIGO and BBO, as shown in figure \ref{fig:ads_ecc_339}. Each subplot represents the case of an SS-SP system with different orbital eccentricities, i.e., a circular orbit with $e = 0$, a mildly eccentric orbit with $e = 0.5$, and a highly eccentric orbit with $e = 0.95$. Each filled circle in the plot represents an \(\text{ASD}_n\) corresponding to the $n^{th}$ harmonic. When the orbit is circular, the dominant signal appears when harmonic $n = 2$. However, when the orbit is eccentric, the signals appear at different harmonic frequencies with different strengths. As can be observed from this figure, the effects of planet mass $m_2$ and distance $D_L$ on the signal strength are significant. From top to bottom, the strength of the signal increases with the increase of planet mass $m_2$, while from left to right, it also increases with the decrease of distance $D_L$. For example, as shown in figure \ref{fig:ads_ecc_339}(a), the signal is below the sensitivity curves of both DECIGO and BBO when the planet mass $m_2 = 10^{-7} \, {\rm M_{\odot}}$ and the distance $D_L = 10 \, \text{kpc}$. In contrast, the signals are strong enough and lie above the sensitivity curves of both DECIGO and BBO when the planet mass $m_2 = 10^{-3} \, {\rm M_{\odot}}$ and the distance $D_L = 0.1 \, \text{kpc}$, as shown in figure \ref{fig:ads_ecc_339}(i).

The \(\text{S/N}\) values calculated for the discussed cases in figure \ref{fig:ads_ecc_339} are listed in the upper part of table \ref{tab:table_2}. We set \(\text{S/N} = 5\) as the detectable threshold following previous works \citep[e.g.,][]{Kupfer2018MNRAS,Kurban2026AA}. Systems with \(\text{S/N}\) values exceeding this threshold are considered detectable. As can be seen from the table and figure, some configurations are detectable by DECIGO and BBO with slightly different $\text{S/N}$ values, due to the sensitivity difference between the two detectors. In the case of circular orbits, the configurations corresponding to figure \ref{fig:ads_ecc_339}(f), (g), (h), and (i) can be detected by both DECIGO and BBO with a $\text{S/N}\geq5$. For eccentric orbits, the $\text{S/N}$ values exhibit an increase compared to the corresponding cases in circular orbits. When the eccentricity $e = 0.5$, the configurations capable of generating a $\text{S/N}$ larger than 5 correspond to figure \ref{fig:ads_ecc_339}(e), (f), (g), (h), and (i) for both DECIGO and BBO. When $e = 0.95$, the configurations for which DECIGO fails to achieve a $\text{S/N}$ less than 5 are figure \ref{fig:ads_ecc_339}(a), (b), and for BBO, it is figure \ref{fig:ads_ecc_339}(a). These findings suggest that eccentricity can significantly enhance the signal strength and render detectability feasible. 

To demonstrate the influence of the primary mass $m_1$ on the results, calculations are carried out for $m_1 = 2 \, {\rm M_{\odot}}$, while other parameters such as $m_2$, $f_{\rm orb}$, $D_L$, and $e$ are maintained at the same values as in figure \ref{fig:ads_ecc_339}. The results are presented in the lower part of table \ref{tab:table_2}. Evidently, from the table, the $\text{S/N}$ values of all cases exhibit an approximate 27\% increase compared to the case of $m_1 = 1.4 \, {\rm M_{\odot}}$, suggesting that an increment in $m_1$ can enhance the signal considerably.

\subsection{Constraints on the detectable parameter space} \label{subsec:detectable_para}

To illustrate the potential scope of the detectable parameter space, calculations are conducted for a fixed primary mass \(m_1 = 1.4 \, {\rm M_{\odot}}\), eccentricity \(e = \{0, 0.5, 0.95\}\), and distance \(D_L = \{0.1, 10, 1000\} \, \text{kpc}\). This enables us to constrain the possible ranges of the planet mass \(m_2\), the orbital frequency \(f_{\rm orb}\), and the \(\text{S/N}\) for each configuration. The \(\text{S/N}\) values of each configuration are obtained using the same method as in Section \ref{subsec:asd and snr}.

The parameter space in which the SS-SP systems are detectable by DECIGO with $\text{S/N} \geq 5$ is shown in figure \ref{fig:fig_para_snr_decigo}. Each subplot of this figure represents a case with a different group of parameter sets. From this figure, it can be observed that the ranges of $m_2$, $f_{\rm orb}$, and $\text{S/N}$ vary among different parameter sets. For a fixed eccentricity $e$, the lower bounds of the orbital frequencies $f_{\rm orb}^{\rm min}$ and planet masses $m_{2}^{\rm min}$ that are detectable by DECIGO increase as the distance increases. For example, when $e = 0$, $f_{\rm orb}^{\rm min} = \{8.6\times10^{-4}, 5.3\times10^{-3}, 3.1\times10^{-2}\} \, \text{Hz}$ and $m_{2}^{\rm min} = \{7.6\times10^{-10}, 7.6\times10^{-8}, 7.2\times10^{-6}\} \, {\rm M_{\odot}}$ for $D_L = \{0.1, 10, 1000\} \, \text{kpc}$. 
For a fixed distance $D_L$, an increase in eccentricity $e$ causes the detectable $f_{\rm orb}$ to shift towards lower frequency bands in general, which further results in an increase in the detectable mass $m_2$. For example, when $D_L = 1000 \, \text{kpc}$, $f_{\rm orb}^{\rm max} = \{6.6, 2.3, 7.4\times10^{-2}\} \, \text{Hz}$ and $m_{2}^{\rm min} = \{7.2\times10^{-6}, 9.5\times10^{-6}, 4.7\times10^{-5}\} \, {\rm M_{\odot}}$ for $e = \{0, 0.5, 0.95\}$.
It can be clearly observed that the degree of contraction becomes more significant as both distance and eccentricity increase. For instance, when $\{e, D_L\} = \{0, 0.1 \, \text{kpc}\}$, the minimum and maximum orbital frequencies $\{f_{\rm orb}^{\rm min}, f_{\rm orb}^{\rm max}\} = \{8.6\times10^{-4}, 6.6\} \,\text{Hz}$, and $m_{2}^{\text{min}} = 7.6\times10^{-10} \, {\rm M_{\odot}}$. Similarly, for more eccentric and distant sources with $\{e, D_L\} = \{0.95, 1000 \, \text{kpc}\}$, $\{f_{\rm orb}^{\rm min}, f_{\rm orb}^{\rm max}\} = \{1.1\times10^{-3}, 7.4\times10^{-2}\} \,\text{Hz}$, and $m_{2}^{\text{min}} = 4.7\times10^{-5} \, {\rm M_{\odot}}$. For each configuration, the maximum S/N is achieved when \( m_{2} = 10^{-3} \, {\rm M_{\odot}} \). Specifically, the maximum S/N values are \( 5.4\times10^{6} \) and \( 206 \), which correspond to the parameter sets \( \{e, D_L\} = \{0, 0.1 \, \text{kpc}\} \) and \( \{e, D_L\} = \{0.95, 1000 \, \text{kpc}\} \), respectively.

The computational results for BBO obtained using the same parameter set employed for the DECIGO case are presented in figure \ref{fig:fig_para_snr_bbo}. As illustrated in the figure, the characteristics of the detectable mass and orbital frequency follow the same overall trend as in the DECIGO case (figure \ref{fig:fig_para_snr_decigo}), with only minor deviations. For example, when \( \{e, D_L\} = \{0, 0.1 \, \text{kpc}\} \), we obtain \(f_{\rm orb}^{\rm min} = 7.7\times10^{-4} \,\text{Hz}\) and \(m_{2}^{\rm min} = 4.3\times10^{-10} \, {\rm M_{\odot}}\); when \( \{e, D_L\} = \{0.95, 1000 \, \text{kpc}\} \), we get \(f_{\rm orb}^{\rm min} = 5.9\times10^{-4} \,\text{Hz}\) and \(m_{2}^{\rm min} = 2.3\times10^{-5} \, {\rm M_{\odot}}\). It can be seen from this BBO configuration that both $f_{\rm orb}^{\rm min}$ and $m_{2}^{\rm min}$ are slightly lower than the corresponding values obtained for DECIGO, and the magnitude of the difference varies across different configurations. This pattern arises primarily because BBO exhibits higher sensitivity than DECIGO over the GW frequency band ranging from roughly \(7\times10^{-2} \,\text{Hz}\) to \(9\times10^{-1} \,\text{Hz}\), which is illustrated in figure \ref{fig:ads_ecc_339}.

\section{Conclusions and Discussion} \label{sec:conclusion_discussion}

In this study, we have investigated the characteristics of the GWs produced from SS-SP systems using space-based GW detectors DECIGO and BBO. Attention is paid to physically viable parameter space for calculating their GW emission properties. We compared the GW signals from our model systems with four years of observation time to the sensitivity curves of DECIGO and BBO in the amplitude spectral density versus GW frequency diagram and calculated their S/N values. We also constrained the parameter space within which SS-SP systems can be detected by DECIGO and BBO. The main results are summarised below.

Small celestial bodies composed of SQM that form around strange stars can be detected by next-generation space-based GW observatories such as DECIGO and BBO. The results of this study demonstrate that both signal amplitude and the corresponding S/N values are influenced by the parameters of the system, including the mass of the host strange star, the mass of the strange planet, orbital separation (orbital frequency), distance to the system, and orbital eccentricity. For a fixed host star mass \(m_1 = 1.4 \, {\rm M_{\odot}}\) and an observation duration \(T_{\rm obs} = 4\) years, our calculation results indicate that the parameter space in which SS-SP systems can be detected with a S/N ratio greater than or equal to 5 is extensive. This parameter space shrinks as the distance to the system and orbital eccentricity increase. For orbital and distance parameters \(\{e, D_L\} = \{0, 0.1 \, \text{kpc}\}\), DECIGO can detect small bodies with masses \( \geq 7.6\times10^{-10} \, {\rm M_{\odot}}\); for \(\{e, D_L\} = \{0.95, 1000 \, \text{kpc}\}\), the detectable mass is \( \geq 4.7\times10^{-5} \, {\rm M_{\odot}}\). Orbital eccentricity enables the detection of systems characterized by long orbital periods. For BBO, the orders of magnitude of the detectable parameter space for orbital frequency and planetary mass are generally consistent across the two detectors. However, within its most sensitive frequency band, BBO can detect celestial sources with minimum planet masses equal to half the lower limit of masses detectable by DECIGO.

The detection or non-detection of SS-SP systems has important implications for compact star physics. The detection of such systems with various parameters could not only provide important clues for the existence of SQM objects, but also offer important information about the formation mechanisms of these systems and related explosive events. The non-detection of SS-SP systems would not exclude the existence of SQM objects, because the possibility of detecting strange stars produced by the hadron-quark phase transition in binary NS mergers \citep[e.g.,][]{2019PhRvL.122f1102B,2020PhRvL.124q1103W,2020PhRvL.125n1103B,2026PhRvD.113d4057H} or in core-collapse supernovae \citep{2020PhRvL.125e1102Z} through identifying the GW signatures at high frequency range still exists.

How strange planets form is an interesting question to answer. The formation of SQM planets can be attributed to the following processes. On the one hand, the birth of a strange star formed from fierce explosive events may lead to the formation of a strange planet. First, a large amount of SQM nuggets may be ejected from the newborn strange star, which can contaminate the surrounding normal planets and convert them into strange planets \citep{1987PhLB..192...71O,2014PhLB..733..164P}. 
Second, a large clump of SQM may be directly ejected from a newborn hot strange star due to the joint effects of fast spinning and high turbulence, forming a strange planet outside \citep{1998ChPhL..15..934X,2006APh....25..212X,2012RAA....12..813H}. Recent analyses indicate that larger SQM clumps formed from these explosive events could survive \citep{2022PhRvD.106j3032B,2025PhRvL.135i1402M}. On the other hand, the planetary-mass SQM clumps may be formed at the quark-gluon phase of the early Universe \citep{Witten1984PhRvD}, which can survive to the present \citep{1994PhRvL..73.1328C,2026arXiv260316686Q}. They can also be captured by compact stars, forming planetary systems. 

It is worth noting that the S/N values we calculated were obtained by averaging S/N across the source's sky position, orbital inclination, and polarization angle. In practice, these geometrical parameters of an SS-SP system can change the S/N by a factor of several. This variation originates from the amplitude modulation induced by the detector's orbital motion with respect to the GW source \citep{Huang2020PhRvD,Wagg2022ApJS}. In addition, extending the observation duration conducted by independent detectors or using the DECIGO and BBO for network observation can considerably enhance the S/N \citep[e.g.,][]{Kurban2026AA}. The improvement of the S/N achieved by incorporating these factors makes it possible to detect planets with masses smaller than claimed above.

In short, the results of this study demonstrate that the next-generation space-based GW detectors DECIGO and BBO hold the potential to detect GWs from SS-SP systems within the Milky Way and nearby galaxies within a distance \(\lesssim 1000 \, \text{kpc}\). Observing these systems via GW astronomy offers an alternative method to test the SQM hypothesis, and is of great importance for research in both astrophysics and fundamental physics.

\section{Acknowledgments}

This study was supported by the National Natural Science Foundation of China (Grant Nos. 12288102, 12573051, 12233002),
the Xinjiang Talent Development Fund No. XJRC-2025-KJ-PY-KJLJ-106, the National Key R\&D Program of China (Nos. 2021YFA0718500, 2022YFA1603104),
the Tianshan Talent Training Program of Xinjiang Uygur Autonomous Region, China (Grant Nos. 2023TSYCLJ0053, 2023TSYCTD0013).
Y.F.H. acknowledges the support from the Xinjiang Tianchi Program.
A.K. acknowledges the support from the Tianchi Talents Project of Xinjiang Uygur Autonomous Region.
WMY is supported by the Tianshan Talent Training Program (No. 2024TSYCCX0073) and the CAS project (No. JZHKYPT-2021-06).
JHC acknowledges the National Natural Science Foundation of China (the NSFC Type C Youth Project 12503053).
R.Y. is supported by the National Key Program for Science and  Technology Research and Development No. 2022YFC2205201, the National SKA  Program of China No. 2020SKA0120200, the National Natural Science Foundation of China (NSFC) project (Nos. 12041303, 12041304), the Major Science and Technology Program of Xinjiang Uygur Autonomous Region No. 2022A03013-2, 2022A03013-4. 
Y.B.W. acknowledges the support from the National Natural Science Foundation of China (No. 12403054), the Sichuan Provincial Natural Science Foundation Project (No. 2025ZNSFSC0878).
Z.G.W. is supported by the Major Science and Technology Program of Xinjiang Uygur Autonomous Region (No. 2022A03013-1), the National Natural Science Foundation of China (No. 12303053), the Youth Innovation Promotion Association of CAS (No. 2023069), and the Tianshan Talent Training Program (No. 2023TSYCCX0100).
This research is partly supported by the Operation, Maintenance and Upgrading Fund for Astronomical Telescopes and Facility Instruments, budgeted from the Ministry of Finance of China (MOF) and administrated by the CAS.


\nocite{*}
\bibliographystyle{aasjournal}
\bibliography{reference}

@ARTICLE{Farhi1984PhRvD,
	author = {{Farhi}, Edward and {Jaffe}, R.~L.},
	title = "{Strange matter}",
	journal = {\prd},
	year = 1984,
	month = dec,
	volume = {30},
	number = {11},
	pages = {2379-2390},
	doi = {10.1103/PhysRevD.30.2379},
	adsurl = {https://ui.adsabs.harvard.edu/abs/1984PhRvD..30.2379F}
}

@ARTICLE{Bodmer1971,
	author = {{Bodmer}, A.~R.},
	journal = {Phys. Rev. D},
	year = 1971,
	month = sep,
	volume = {4},
	number = {6},
	pages = {1601-1606},
	doi = {10.1103/PhysRevD.4.1601},
	adsurl = {https://ui.adsabs.harvard.edu/abs/1971PhRvD...4.1601B}
}

@ARTICLE{Witten1984PhRvD,
	author = {{Witten}, Edward},
	title = "{Cosmic separation of phases}",
	journal = {\prd},
	year = 1984,
	month = jul,
	volume = {30},
	number = {2},
	pages = {272-285},
	doi = {10.1103/PhysRevD.30.272},
	adsurl = {https://ui.adsabs.harvard.edu/abs/1984PhRvD..30..272W}
}

@ARTICLE{Haensel1986AA,
	author = {{Haensel}, P. and {Zdunik}, J.~L. and {Schaefer}, R.},
	title = "{Strange quark stars}",
	journal = {\aap},
	year = 1986,
	month = may,
	volume = {160},
	number = {1},
	pages = {121-128},
	adsurl = {https://ui.adsabs.harvard.edu/abs/1986A&A...160..121H}
}

@ARTICLE{Kettner1995PhRvD,
	author = {{Kettner}, Ch. and {Weber}, F. and {Weigel}, M.~K. and {Glendenning}, N.~K.},
	title = "{Structure and stability of strange and charm stars at finite temperatures}",
	journal = {\prd},
	year = 1995,
	month = feb,
	volume = {51},
	number = {4},
	pages = {1440-1457},
	doi = {10.1103/PhysRevD.51.1440},
	adsurl = {https://ui.adsabs.harvard.edu/abs/1995PhRvD..51.1440K}
}

@ARTICLE{2020PhRvL.124q1103W,
	author = {{Weih}, Lukas R. and {Hanauske}, Matthias and {Rezzolla}, Luciano},
	title = "{Postmerger Gravitational-Wave Signatures of Phase Transitions in Binary Mergers}",
	journal = {\prl},
	year = 2020,
	month = may,
	volume = {124},
	number = {17},
	eid = {171103},
	pages = {171103},
	doi = {10.1103/PhysRevLett.124.171103},
	archivePrefix = {arXiv},
	eprint = {1912.09340},
	primaryClass = {gr-qc},
	adsurl = {https://ui.adsabs.harvard.edu/abs/2020PhRvL.124q1103W}
}

@ARTICLE{2020PhRvL.125n1103B,
	author = {{Bauswein}, Andreas and {Blacker}, Sebastian and {Vijayan}, Vimal and {Stergioulas}, Nikolaos and {Chatziioannou}, Katerina and {Clark}, James A. and {Bastian}, Niels-Uwe F. and {Blaschke}, David B. and {Cierniak}, Mateusz and {Fischer}, Tobias},
	title = "{Equation of State Constraints from the Threshold Binary Mass for Prompt Collapse of Neutron Star Mergers}",
	journal = {\prl},
	year = 2020,
	month = oct,
	volume = {125},
	number = {14},
	eid = {141103},
	pages = {141103},
	doi = {10.1103/PhysRevLett.125.141103},
	archivePrefix = {arXiv},
	eprint = {2004.00846},
	primaryClass = {astro-ph.HE},
	adsurl = {https://ui.adsabs.harvard.edu/abs/2020PhRvL.125n1103B}
}

@ARTICLE{2026PhRvD.113d4057H,
	author = {{Hammond}, P. and {Clevinger}, A. and {Albino}, M. and {Dexheimer}, V. and {Bernuzzi}, S. and {Brown}, C. and {Cook}, W. and {Daszuta}, B. and {Fields}, J. and {Grundy}, E. and {Provid{\^e}ncia}, C. and {Radice}, D. and {Steiner}, A.},
	title = "{Investigating the impact of higher-order phase transitions in binary neutron-star mergers}",
	journal = {\prd},
	year = 2026,
	month = feb,
	volume = {113},
	number = {4},
	eid = {044057},
	pages = {044057},
	doi = {10.1103/71t3-3t28},
	archivePrefix = {arXiv},
	eprint = {2508.10698},
	primaryClass = {astro-ph.HE},
	adsurl = {https://ui.adsabs.harvard.edu/abs/2026PhRvD.113d4057H}
}

@ARTICLE{Alcock1986ApJ,
	author = {{Alcock}, Charles and {Farhi}, Edward and {Olinto}, Angela},
	title = "{Strange Stars}",
	journal = {\apj},
	year = 1986,
	month = nov,
	volume = {310},
	pages = {261},
	doi = {10.1086/164679},
	adsurl = {https://ui.adsabs.harvard.edu/abs/1986ApJ...310..261A}
}

@ARTICLE{Geng2015ApJ_a,
	author = {{Geng}, J.~J. and {Huang}, Y.~F. and {Lu}, T.},
	title = "{Coalescence of Strange-quark Planets with Strange Stars: a New Kind of Source for Gravitational Wave Bursts}",
	journal = {\apj},
	year = 2015,
	month = may,
	volume = {804},
	number = {1},
	eid = {21},
	pages = {21},
	doi = {10.1088/0004-637X/804/1/21},
	archivePrefix = {arXiv},
	eprint = {1501.02122},
	primaryClass = {astro-ph.HE},
	adsurl = {https://ui.adsabs.harvard.edu/abs/2015ApJ...804...21G}
}

@ARTICLE{Geng2021Innov,
	author = {{Geng}, Jinjun and {Li}, Bing and {Huang}, Yongfeng},
	title = "{Repeating fast radio bursts from collapses of the crust of a strange star}",
	journal = {The Innovation},
	year = 2021,
	month = nov,
	volume = {2},
	eid = {100152},
	pages = {100152},
	doi = {10.1016/j.xinn.2021.100152},
	archivePrefix = {arXiv},
	eprint = {2103.04165},
	primaryClass = {astro-ph.HE},
	adsurl = {https://ui.adsabs.harvard.edu/abs/2021Innov...200152G}
}

@ARTICLE{Huang2017ApJ,
	author = {{Huang}, Y.~F. and {Yu}, Y.~B.},
	title = "{Searching for Strange Quark Matter Objects in Exoplanets}",
	journal = {\apj},
	year = 2017,
	month = oct,
	volume = {848},
	number = {2},
	eid = {115},
	pages = {115},
	doi = {10.3847/1538-4357/aa8b63},
	archivePrefix = {arXiv},
	eprint = {1702.07978},
	primaryClass = {astro-ph.HE},
	adsurl = {https://ui.adsabs.harvard.edu/abs/2017ApJ...848..115H}
}

@INPROCEEDINGS{Kuerban2019AIPC,
	author = {{Kuerban}, Abudushataer and {Geng}, Jin-Jun and {Huang}, Yong-Feng},
	title = "{GW emission from merging strange quark star-strange quark planet systems}",
	booktitle = {Xiamen-CUSTIPEN Workshop on the Equation of State of Dense Neutron-Rich Matter in the Era of Gravitational Wave Astronomy},
	year = 2019,
	series = {American Institute of Physics Conference Series},
	volume = {2127},
	month = jul,
	eid = {020027},
	pages = {020027},
	doi = {10.1063/1.5117817},
	adsurl = {https://ui.adsabs.harvard.edu/abs/2019AIPC.2127b0027K}
}

@ARTICLE{Kuerban2020ApJ,
	author = {{Kuerban}, Abudushataer and {Geng}, Jin-Jun and {Huang}, Yong-Feng and
	{Zong}, Hong-Shi and {Gong}, Hang},
	title = "{Close-in Exoplanets as Candidates for Strange Quark Matter Objects}",
	journal = {\apj},
	year = 2020,
	month = feb,
	volume = {890},
	number = {1},
	eid = {41},
	pages = {41},
	doi = {10.3847/1538-4357/ab698b},
	archivePrefix = {arXiv},
	eprint = {1908.11191},
	primaryClass = {astro-ph.HE},
	adsurl = {https://ui.adsabs.harvard.edu/abs/2020ApJ...890...41K}
}

@ARTICLE{2022PhLB..83237204K,
	author = {{Kurban}, Abdusattar and {Huang}, Yong-Feng and {Geng}, Jin-Jun and {Zong}, Hong-Shi},
	title = "{Searching for strange quark matter objects among white dwarfs}",
	journal = {Physics Letters B},
	year = 2022,
	month = sep,
	volume = {832},
	eid = {137204},
	pages = {137204},
	doi = {10.1016/j.physletb.2022.137204},
	adsurl = {https://ui.adsabs.harvard.edu/abs/2022PhLB..83237204K}
}

@ARTICLE{2022ApJ...928...94K,
	author = {{Kurban}, Abdusattar and {Huang}, Yong-Feng and {Geng}, Jin-Jun and {Li}, Bing and {Xu}, Fan and {Wang}, Xu and {Zhou}, Xia and {Esamdin}, Ali and {Wang}, Na},
	title = "{Periodic Repeating Fast Radio Bursts: Interaction between a Magnetized Neutron Star and Its Planet in an Eccentric Orbit}",
	journal = {\apj},
	year = 2022,
	month = mar,
	volume = {928},
	number = {1},
	eid = {94},
	pages = {94},
	doi = {10.3847/1538-4357/ac558f},
	archivePrefix = {arXiv},
	eprint = {2102.04264},
	primaryClass = {astro-ph.HE},
	adsurl = {https://ui.adsabs.harvard.edu/abs/2022ApJ...928...94K}
}

@ARTICLE{2023MNRAS.522.4265K,
	author = {{Kurban}, Abdusattar and {Zhou}, Xia and {Wang}, Na and {Huang}, Yong-Feng and {Wang}, Yu-Bin and {Nurmamat}, Nurimangul},
	title = "{Dynamics of the clumps partially disrupted from a planet around a neutron star}",
	journal = {\mnras},
	year = 2023,
	month = jul,
	volume = {522},
	number = {3},
	pages = {4265-4274},
	doi = {10.1093/mnras/stad1260},
	archivePrefix = {arXiv},
	eprint = {2305.08142},
	primaryClass = {astro-ph.HE},
	adsurl = {https://ui.adsabs.harvard.edu/abs/2023MNRAS.522.4265K}
}

@ARTICLE{2024A&A...686A..87K,
	author = {{Kurban}, Abdusattar and {Zhou}, Xia and {Wang}, Na and {Huang}, Yong-Feng and {Wang}, Yu-Bin and {Nurmamat}, Nurimangul},
	title = "{Repeating X-ray bursts: Interaction between a neutron star and clumps partially disrupted from a planet}",
	journal = {\aap},
	year = 2024,
	month = jun,
	volume = {686},
	eid = {A87},
	pages = {A87},
	doi = {10.1051/0004-6361/202347828},
	archivePrefix = {arXiv},
	eprint = {2403.13333},
	primaryClass = {astro-ph.HE},
	adsurl = {https://ui.adsabs.harvard.edu/abs/2024A&A...686A..87K}
}

@ARTICLE{2024EPJC...84..210N,
	author = {{Nurmamat}, Nurimangul and {Huang}, Yong-Feng and {Geng}, Jin-Jun and {Kurban}, Abdusattar and {Li}, Bing},
	title = "{Repeating fast radio bursts produced by a strange star interacting with its planet in an eccentric orbit}",
	journal = {European Physical Journal C},
	year = 2024,
	month = feb,
	volume = {84},
	number = {2},
	eid = {210},
	pages = {210},
	doi = {10.1140/epjc/s10052-024-12572-5},
	archivePrefix = {arXiv},
	eprint = {2211.12026},
	primaryClass = {astro-ph.HE},
	adsurl = {https://ui.adsabs.harvard.edu/abs/2024EPJC...84..210N}
}

@ARTICLE{2024FrASS..1109463Z,
	author = {{Zhang}, Xiao-Li and {Huang}, Yong-Feng and {Zou}, Ze-Cheng},
	title = "{Recent progresses in strange quark stars}",
	journal = {Frontiers in Astronomy and Space Sciences},
	year = 2024,
	month = aug,
	volume = {11},
	eid = {1409463},
	pages = {1409463},
	doi = {10.3389/fspas.2024.1409463},
	archivePrefix = {arXiv},
	eprint = {2404.00363},
	primaryClass = {astro-ph.HE},
	adsurl = {https://ui.adsabs.harvard.edu/abs/2024FrASS..1109463Z}
}

@ARTICLE{Zhang2024MNRAS,
	author = {{Zhang}, Xiao-Li and {Zou}, Ze-Cheng and {Huang}, Yong-Feng and {Gao}, Hao-Xuan and {Wang}, Pei and {Cui}, Lang and {Liu}, Xiang},
	title = "{Gravitational wave emission from close-in strange quark planets around strange stars with magnetic interactions}",
	journal = {\mnras},
	year = 2024,
	month = jul,
	volume = {531},
	number = {4},
	pages = {3905-3911},
	doi = {10.1093/mnras/stae1400},
	archivePrefix = {arXiv},
	eprint = {2402.00730},
	primaryClass = {astro-ph.HE},
	adsurl = {https://ui.adsabs.harvard.edu/abs/2024MNRAS.531.3905Z}
}

@ARTICLE{2016PhRvL.116f1102A,
	author = {{Abbott}, B.~P. and {Abbott}, R. and {Abbott}, T.~D. and {Abernathy}, M.~R. and {Acernese}, F. and {Ackley}, K. and {Adams}, C. and {Adams}, T. and {Addesso}, P. and {Adhikari}, R.~X. and {Adya}, V.~B. and {Affeldt}, C. and {Agathos}, M. and {Agatsuma}, K. and {Aggarwal}, N. and {Aguiar}, O.~D. and {Aiello}, L. and {Ain}, A. and {Ajith}, P. and {Allen}, B. and {Allocca}, A. and {Altin}, P.~A. and {Anderson}, S.~B. and {Anderson}, W.~G. and {Arai}, K. and {Arain}, M.~A. and {Araya}, M.~C. and {Arceneaux}, C.~C. and {Areeda}, J.~S. and {Arnaud}, N. and {Arun}, K.~G. and {Ascenzi}, S. and {Ashton}, G. and {Ast}, M. and {Aston}, S.~M. and {Astone}, P. and {Aufmuth}, P. and {Aulbert}, C. and {Babak}, S. and {Bacon}, P. and {Bader}, M.~K.~M. and {Baker}, P.~T. and {Baldaccini}, F. and {Ballardin}, G. and {Ballmer}, S.~W. and {Barayoga}, J.~C. and {Barclay}, S.~E. and {Barish}, B.~C. and {Barker}, D. and {Barone}, F. and {Barr}, B. and {Barsotti}, L. and {Barsuglia}, M. and {Barta}, D. and {Bartlett}, J. and {Barton}, M.~A. and {Bartos}, I. and {Bassiri}, R. and {Basti}, A. and {Batch}, J.~C. and {Baune}, C. and {Bavigadda}, V. and {Bazzan}, M. and {Behnke}, B. and {Bejger}, M. and {Belczynski}, C. and {Bell}, A.~S. and {Bell}, C.~J. and {Berger}, B.~K. and {Bergman}, J. and {Bergmann}, G. and {Berry}, C.~P.~L. and {Bersanetti}, D. and {Bertolini}, A. and {Betzwieser}, J. and {Bhagwat}, S. and {Bhandare}, R. and {Bilenko}, I.~A. and {Billingsley}, G. and {Birch}, J. and {Birney}, I.~A. and {Birnholtz}, O. and {Biscans}, S. and {Bisht}, A. and {Bitossi}, M. and {Biwer}, C. and {Bizouard}, M.~A. and {Blackburn}, J.~K. and {Blair}, C.~D. and {Blair}, D.~G. and {Blair}, R.~M. and {Bloemen}, S. and {Bock}, O. and {Bodiya}, T.~P. and {Boer}, M. and {Bogaert}, G. and {Bogan}, C. and {Bohe}, A. and {Bojtos}, P. and {Bond}, C. and {Bondu}, F. and {Bonnand}, R. and {Boom}, B.~A. and {Bork}, R. and {Boschi}, V. and {Bose}, S. and {Bouffanais}, Y. and {Bozzi}, A. and {Bradaschia}, C. and {Brady}, P.~R. and {Braginsky}, V.~B. and {Branchesi}, M. and {Brau}, J.~E. and {Briant}, T. and {Brillet}, A. and {Brinkmann}, M. and {Brisson}, V. and {Brockill}, P. and {Brooks}, A.~F. and {Brown}, D.~A. and {Brown}, D.~D. and {Brown}, N.~M. and {Buchanan}, C.~C. and {Buikema}, A. and {Bulik}, T. and {Bulten}, H.~J. and {Buonanno}, A. and {Buskulic}, D. and {Buy}, C. and {Byer}, R.~L. and {Cabero}, M. and {Cadonati}, L. and {Cagnoli}, G. and {Cahillane}, C. and {Bustillo}, J. Calder{\'o}n and {Callister}, T. and {Calloni}, E. and {Camp}, J.~B. and {Cannon}, K.~C. and {Cao}, J. and {Capano}, C.~D. and {Capocasa}, E. and {Carbognani}, F. and {Caride}, S. and {Diaz}, J. Casanueva and {Casentini}, C. and {Caudill}, S. and {Cavagli{\`a}}, M. and {Cavalier}, F. and {Cavalieri}, R. and {Cella}, G. and {Cepeda}, C.~B. and {Baiardi}, L. Cerboni and {Cerretani}, G. and {Cesarini}, E. and {Chakraborty}, R. and {Chalermsongsak}, T. and {Chamberlin}, S.~J. and {Chan}, M. and {Chao}, S. and {Charlton}, P. and {Chassande-Mottin}, E. and {Chen}, H.~Y. and {Chen}, Y. and {Cheng}, C. and {Chincarini}, A. and {Chiummo}, A. and {Cho}, H.~S. and {Cho}, M. and {Chow}, J.~H. and {Christensen}, N. and {Chu}, Q. and {Chua}, S. and {Chung}, S. and {Ciani}, G. and {Clara}, F. and {Clark}, J.~A. and {Cleva}, F. and {Coccia}, E. and {Cohadon}, P. -F. and {Colla}, A. and {Collette}, C.~G. and {Cominsky}, L. and {Constancio}, M. and {Conte}, A. and {Conti}, L. and {Cook}, D. and {Corbitt}, T.~R. and {Cornish}, N. and {Corsi}, A. and {Cortese}, S. and {Costa}, C.~A. and {Coughlin}, M.~W. and {Coughlin}, S.~B. and {Coulon}, J. -P. and {Countryman}, S.~T. and {Couvares}, P. and {Cowan}, E.~E. and {Coward}, D.~M. and {Cowart}, M.~J.},
	title = "{Observation of Gravitational Waves from a Binary Black Hole Merger}",
	journal = {\prl},
	year = 2016,
	month = feb,
	volume = {116},
	number = {6},
	eid = {061102},
	pages = {061102},
	doi = {10.1103/PhysRevLett.116.061102},
	archivePrefix = {arXiv},
	eprint = {1602.03837},
	primaryClass = {gr-qc},
	adsurl = {https://ui.adsabs.harvard.edu/abs/2016PhRvL.116f1102A}
}

@ARTICLE{Abbott2017PhRvL.119p1101A,
	author = {{Abbott}, B.~P. and {Abbott}, R. and {Abbott}, T.~D. and {Acernese}, F. and {Ackley}, K. and {Adams}, C. and {Adams}, T. and {Addesso}, P. and {Adhikari}, R.~X. and {Adya}, V.~B. and {Affeldt}, C. and {Afrough}, M. and {Agarwal}, B. and {Agathos}, M. and {Agatsuma}, K. and {Aggarwal}, N. and {Aguiar}, O.~D. and {Aiello}, L. and {Ain}, A. and {Ajith}, P. and {Allen}, B. and {Allen}, G. and {Allocca}, A. and {Altin}, P.~A. and {Amato}, A. and {Ananyeva}, A. and {Anderson}, S.~B. and {Anderson}, W.~G. and {Angelova}, S.~V. and {Antier}, S. and {Appert}, S. and {Arai}, K. and {Araya}, M.~C. and {Areeda}, J.~S. and {Arnaud}, N. and {Arun}, K.~G. and {Ascenzi}, S. and {Ashton}, G. and {Ast}, M. and {Aston}, S.~M. and {Astone}, P. and {Atallah}, D.~V. and {Aufmuth}, P. and {Aulbert}, C. and {AultONeal}, K. and {Austin}, C. and {Avila-Alvarez}, A. and {Babak}, S. and {Bacon}, P. and {Bader}, M.~K.~M. and {Bae}, S. and {Bailes}, M. and {Baker}, P.~T. and {Baldaccini}, F. and {Ballardin}, G. and {Ballmer}, S.~W. and {Banagiri}, S. and {Barayoga}, J.~C. and {Barclay}, S.~E. and {Barish}, B.~C. and {Barker}, D. and {Barkett}, K. and {Barone}, F. and {Barr}, B. and {Barsotti}, L. and {Barsuglia}, M. and {Barta}, D. and {Barthelmy}, S.~D. and {Bartlett}, J. and {Bartos}, I. and {Bassiri}, R. and {Basti}, A. and {Batch}, J.~C. and {Bawaj}, M. and {Bayley}, J.~C. and {Bazzan}, M. and {B{\'e}csy}, B. and {Beer}, C. and {Bejger}, M. and {Belahcene}, I. and {Bell}, A.~S. and {Berger}, B.~K. and {Bergmann}, G. and {Bernuzzi}, S. and {Bero}, J.~J. and {Berry}, C.~P.~L. and {Bersanetti}, D. and {Bertolini}, A. and {Betzwieser}, J. and {Bhagwat}, S. and {Bhandare}, R. and {Bilenko}, I.~A. and {Billingsley}, G. and {Billman}, C.~R. and {Birch}, J. and {Birney}, R. and {Birnholtz}, O. and {Biscans}, S. and {Biscoveanu}, S. and {Bisht}, A. and {Bitossi}, M. and {Biwer}, C. and {Bizouard}, M.~A. and {Blackburn}, J.~K. and {Blackman}, J. and {Blair}, C.~D. and {Blair}, D.~G. and {Blair}, R.~M. and {Bloemen}, S. and {Bock}, O. and {Bode}, N. and {Boer}, M. and {Bogaert}, G. and {Bohe}, A. and {Bondu}, F. and {Bonilla}, E. and {Bonnand}, R. and {Boom}, B.~A. and {Bork}, R. and {Boschi}, V. and {Bose}, S. and {Bossie}, K. and {Bouffanais}, Y. and {Bozzi}, A. and {Bradaschia}, C. and {Brady}, P.~R. and {Branchesi}, M. and {Brau}, J.~E. and {Briant}, T. and {Brillet}, A. and {Brinkmann}, M. and {Brisson}, V. and {Brockill}, P. and {Broida}, J.~E. and {Brooks}, A.~F. and {Brown}, D.~A. and {Brown}, D.~D. and {Brunett}, S. and {Buchanan}, C.~C. and {Buikema}, A. and {Bulik}, T. and {Bulten}, H.~J. and {Buonanno}, A. and {Buskulic}, D. and {Buy}, C. and {Byer}, R.~L. and {Cabero}, M. and {Cadonati}, L. and {Cagnoli}, G. and {Cahillane}, C. and {Calder{\'o}n Bustillo}, J. and {Callister}, T.~A. and {Calloni}, E. and {Camp}, J.~B. and {Canepa}, M. and {Canizares}, P. and {Cannon}, K.~C. and {Cao}, H. and {Cao}, J. and {Capano}, C.~D. and {Capocasa}, E. and {Carbognani}, F. and {Caride}, S. and {Carney}, M.~F. and {Carullo}, G. and {Casanueva Diaz}, J. and {Casentini}, C. and {Caudill}, S. and {Cavagli{\`a}}, M. and {Cavalier}, F. and {Cavalieri}, R. and {Cella}, G. and {Cepeda}, C.~B. and {Cerd{\'a}-Dur{\'a}n}, P. and {Cerretani}, G. and {Cesarini}, E. and {Chamberlin}, S.~J. and {Chan}, M. and {Chao}, S. and {Charlton}, P. and {Chase}, E. and {Chassande-Mottin}, E. and {Chatterjee}, D. and {Chatziioannou}, K. and {Cheeseboro}, B.~D. and {Chen}, H.~Y. and {Chen}, X. and {Chen}, Y. and {Cheng}, H. -P. and {Chia}, H. and {Chincarini}, A. and {Chiummo}, A. and {Chmiel}, T. and {Cho}, H.~S. and {Cho}, M. and {Chow}, J.~H. and {Christensen}, N. and {Chu}, Q. and {Chua}, A.~J.~K. and {Chua}, S.},
	title = "{GW170817: Observation of Gravitational Waves from a Binary Neutron Star Inspiral}",
	journal = {\prl},
	year = 2017,
	month = oct,
	volume = {119},
	number = {16},
	eid = {161101},
	pages = {161101},
	doi = {10.1103/PhysRevLett.119.161101},
	archivePrefix = {arXiv},
	eprint = {1710.05832},
	primaryClass = {gr-qc},
	adsurl = {https://ui.adsabs.harvard.edu/abs/2017PhRvL.119p1101A}
}

@ARTICLE{Hills1975Natur,
	author = {{Hills}, J.~G.},
	title = "{Possible power source of Seyfert galaxies and QSOs}",
	journal = {\nat},
	year = 1975,
	month = mar,
	volume = {254},
	number = {5498},
	pages = {295-298},
	doi = {10.1038/254295a0},
	adsurl = {https://ui.adsabs.harvard.edu/abs/1975Natur.254..295H}
}

@ARTICLE{Luo2016CQGra,
	author = {{Luo}, Jun and {Chen}, Li-Sheng and {Duan}, Hui-Zong and {Gong}, Yun-Gui and {Hu}, Shoucun and {Ji}, Jianghui and {Liu}, Qi and {Mei}, Jianwei and {Milyukov}, Vadim and {Sazhin}, Mikhail and {Shao}, Cheng-Gang and {Toth}, Viktor T. and {Tu}, Hai-Bo and {Wang}, Yamin and {Wang}, Yan and {Yeh}, Hsien-Chi and {Zhan}, Ming-Sheng and {Zhang}, Yonghe and {Zharov}, Vladimir and {Zhou}, Ze-Bing},
	title = "{TianQin: a space-borne gravitational wave detector}",
	journal = {Classical and Quantum Gravity},
	year = 2016,
	month = feb,
	volume = {33},
	number = {3},
	eid = {035010},
	pages = {035010},
	doi = {10.1088/0264-9381/33/3/035010},
	archivePrefix = {arXiv},
	eprint = {1512.02076},
	primaryClass = {astro-ph.IM},
	adsurl = {https://ui.adsabs.harvard.edu/abs/2016CQGra..33c5010L}
}

@ARTICLE{Huang2020PhRvD,
	author = {{Huang}, Shun-Jia and {Hu}, Yi-Ming and {Korol}, Valeriya and {Li}, Peng-Cheng and {Liang}, Zheng-Cheng and {Lu}, Yang and {Wang}, Hai-Tian and {Yu}, Shenghua and {Mei}, Jianwei},
	title = "{Science with the TianQin Observatory: Preliminary results on Galactic double white dwarf binaries}",
	journal = {\prd},
	year = 2020,
	month = sep,
	volume = {102},
	number = {6},
	eid = {063021},
	pages = {063021},
	doi = {10.1103/PhysRevD.102.063021},
	archivePrefix = {arXiv},
	eprint = {2005.07889},
	primaryClass = {astro-ph.HE},
	adsurl = {https://ui.adsabs.harvard.edu/abs/2020PhRvD.102f3021H}
}

@ARTICLE{Amaro-Seoane2017arXiv170200786A,
	author = {{Amaro-Seoane}, Pau and {Audley}, Heather and {Babak}, Stanislav and {Baker}, John and {Barausse}, Enrico and {Bender}, Peter and {Berti}, Emanuele and {Binetruy}, Pierre and {Born}, Michael and {Bortoluzzi}, Daniele and {Camp}, Jordan and {Caprini}, Chiara and {Cardoso}, Vitor and {Colpi}, Monica and {Conklin}, John and {Cornish}, Neil and {Cutler}, Curt and {Danzmann}, Karsten and {Dolesi}, Rita and {Ferraioli}, Luigi and {Ferroni}, Valerio and {Fitzsimons}, Ewan and {Gair}, Jonathan and {Gesa Bote}, Lluis and {Giardini}, Domenico and {Gibert}, Ferran and {Grimani}, Catia and {Halloin}, Hubert and {Heinzel}, Gerhard and {Hertog}, Thomas and {Hewitson}, Martin and {Holley-Bockelmann}, Kelly and {Hollington}, Daniel and {Hueller}, Mauro and {Inchauspe}, Henri and {Jetzer}, Philippe and {Karnesis}, Nikos and {Killow}, Christian and {Klein}, Antoine and {Klipstein}, Bill and {Korsakova}, Natalia and {Larson}, Shane L and {Livas}, Jeffrey and {Lloro}, Ivan and {Man}, Nary and {Mance}, Davor and {Martino}, Joseph and {Mateos}, Ignacio and {McKenzie}, Kirk and {McWilliams}, Sean T and {Miller}, Cole and {Mueller}, Guido and {Nardini}, Germano and {Nelemans}, Gijs and {Nofrarias}, Miquel and {Petiteau}, Antoine and {Pivato}, Paolo and {Plagnol}, Eric and {Porter}, Ed and {Reiche}, Jens and {Robertson}, David and {Robertson}, Norna and {Rossi}, Elena and {Russano}, Giuliana and {Schutz}, Bernard and {Sesana}, Alberto and {Shoemaker}, David and {Slutsky}, Jacob and {Sopuerta}, Carlos F. and {Sumner}, Tim and {Tamanini}, Nicola and {Thorpe}, Ira and {Troebs}, Michael and {Vallisneri}, Michele and {Vecchio}, Alberto and {Vetrugno}, Daniele and {Vitale}, Stefano and {Volonteri}, Marta and {Wanner}, Gudrun and {Ward}, Harry and {Wass}, Peter and {Weber}, William and {Ziemer}, John and {Zweifel}, Peter},
	title = "{Laser Interferometer Space Antenna}",
	journal = {arXiv e-prints},
	year = 2017,
	month = feb,
	eid = {arXiv:1702.00786},
	pages = {arXiv:1702.00786},
	doi = {10.48550/arXiv.1702.00786},
	archivePrefix = {arXiv},
	eprint = {1702.00786},
	primaryClass = {astro-ph.IM},
	adsurl = {https://ui.adsabs.harvard.edu/abs/2017arXiv170200786A}
}

@ARTICLE{Amaro-Seoane2023LRR,
	author = {{Amaro-Seoane}, Pau and {Andrews}, Jeff and {Arca Sedda}, Manuel and {Askar}, Abbas and {Baghi}, Quentin and {Balasov}, Razvan and {Bartos}, Imre and {Bavera}, Simone S. and {Bellovary}, Jillian and {Berry}, Christopher P.~L. and {Berti}, Emanuele and {Bianchi}, Stefano and {Blecha}, Laura and {Blondin}, St{\'e}phane and {Bogdanovi{\'c}}, Tamara and {Boissier}, Samuel and {Bonetti}, Matteo and {Bonoli}, Silvia and {Bortolas}, Elisa and {Breivik}, Katelyn and {Capelo}, Pedro R. and {Caramete}, Laurentiu and {Cattorini}, Federico and {Charisi}, Maria and {Chaty}, Sylvain and {Chen}, Xian and {Chru{\'s}li{\'n}ska}, Martyna and {Chua}, Alvin J.~K. and {Church}, Ross and {Colpi}, Monica and {D'Orazio}, Daniel and {Danielski}, Camilla and {Davies}, Melvyn B. and {Dayal}, Pratika and {De Rosa}, Alessandra and {Derdzinski}, Andrea and {Destounis}, Kyriakos and {Dotti}, Massimo and {Du{\c{t}}an}, Ioana and {Dvorkin}, Irina and {Fabj}, Gaia and {Foglizzo}, Thierry and {Ford}, Saavik and {Fouvry}, Jean-Baptiste and {Franchini}, Alessia and {Fragos}, Tassos and {Fryer}, Chris and {Gaspari}, Massimo and {Gerosa}, Davide and {Graziani}, Luca and {Groot}, Paul and {Habouzit}, Melanie and {Haggard}, Daryl and {Haiman}, Zoltan and {Han}, Wen-Biao and {Istrate}, Alina and {Johansson}, Peter H. and {Khan}, Fazeel Mahmood and {Kimpson}, Tomas and {Kokkotas}, Kostas and {Kong}, Albert and {Korol}, Valeriya and {Kremer}, Kyle and {Kupfer}, Thomas and {Lamberts}, Astrid and {Larson}, Shane and {Lau}, Mike and {Liu}, Dongliang and {Lloyd-Ronning}, Nicole and {Lodato}, Giuseppe and {Lupi}, Alessandro and {Ma}, Chung-Pei and {Maccarone}, Tomas and {Mandel}, Ilya and {Mangiagli}, Alberto and {Mapelli}, Michela and {Mathis}, St{\'e}phane and {Mayer}, Lucio and {McGee}, Sean and {McKernan}, Berry and {Miller}, M. Coleman and {Mota}, David F. and {Mumpower}, Matthew and {Nasim}, Syeda S. and {Nelemans}, Gijs and {Noble}, Scott and {Pacucci}, Fabio and {Panessa}, Francesca and {Paschalidis}, Vasileios and {Pfister}, Hugo and {Porquet}, Delphine and {Quenby}, John and {Ricarte}, Angelo and {R{\"o}pke}, Friedrich K. and {Regan}, John and {Rosswog}, Stephan and {Ruiter}, Ashley and {Ruiz}, Milton and {Runnoe}, Jessie and {Schneider}, Raffaella and {Schnittman}, Jeremy and {Secunda}, Amy and {Sesana}, Alberto and {Seto}, Naoki and {Shao}, Lijing and {Shapiro}, Stuart and {Sopuerta}, Carlos and {Stone}, Nicholas C. and {Suvorov}, Arthur and {Tamanini}, Nicola and {Tamfal}, Tomas and {Tauris}, Thomas and {Temmink}, Karel and {Tomsick}, John and {Toonen}, Silvia and {Torres-Orjuela}, Alejandro and {Toscani}, Martina and {Tsokaros}, Antonios and {Unal}, Caner and {V{\'a}zquez-Aceves}, Ver{\'o}nica and {Valiante}, Rosa and {van Putten}, Maurice and {van Roestel}, Jan and {Vignali}, Christian and {Volonteri}, Marta and {Wu}, Kinwah and {Younsi}, Ziri and {Yu}, Shenghua and {Zane}, Silvia and {Zwick}, Lorenz and {Antonini}, Fabio and {Baibhav}, Vishal and {Barausse}, Enrico and {Bonilla Rivera}, Alexander and {Branchesi}, Marica and {Branduardi-Raymont}, Graziella and {Burdge}, Kevin and {Chakraborty}, Srija and {Cuadra}, Jorge and {Dage}, Kristen and {Davis}, Benjamin and {de Mink}, Selma E. and {Decarli}, Roberto and {Doneva}, Daniela and {Escoffier}, Stephanie and {Gandhi}, Poshak and {Haardt}, Francesco and {Lousto}, Carlos O. and {Nissanke}, Samaya and {Nordhaus}, Jason and {O'Shaughnessy}, Richard and {Portegies Zwart}, Simon and {Pound}, Adam and {Schussler}, Fabian and {Sergijenko}, Olga and {Spallicci}, Alessandro and {Vernieri}, Daniele and {Vigna-G{\'o}mez}, Alejandro},
	title = "{Astrophysics with the Laser Interferometer Space Antenna}",
	journal = {Living Reviews in Relativity},
	year = 2023,
	month = dec,
	volume = {26},
	number = {1},
	eid = {2},
	pages = {2},
	doi = {10.1007/s41114-022-00041-y},
	archivePrefix = {arXiv},
	eprint = {2203.06016},
	primaryClass = {gr-qc},
	adsurl = {https://ui.adsabs.harvard.edu/abs/2023LRR....26....2A}
}

@ARTICLE{Ruan2018arXiv180709495R,
	author = {{Ruan}, Wen-Hong and {Guo}, Zong-Kuan and {Cai}, Rong-Gen and {Zhang}, Yuan-Zhong},
	title = "{Taiji Program: Gravitational-Wave Sources}",
	journal = {arXiv e-prints},
	year = 2018,
	month = jul,
	eid = {arXiv:1807.09495},
	pages = {arXiv:1807.09495},
	doi = {10.48550/arXiv.1807.09495},
	archivePrefix = {arXiv},
	eprint = {1807.09495},
	primaryClass = {gr-qc},
	adsurl = {https://ui.adsabs.harvard.edu/abs/2018arXiv180709495R}
}

@ARTICLE{Ruan2020IJMPA,
	author = {{Ruan}, Wen-Hong and {Guo}, Zong-Kuan and {Cai}, Rong-Gen and {Zhang}, Yuan-Zhong},
	title = "{Taiji program: Gravitational-wave sources}",
	journal = {International Journal of Modern Physics A},
	year = 2020,
	month = jun,
	volume = {35},
	number = {17},
	eid = {2050075},
	pages = {2050075},
	doi = {10.1142/S0217751X2050075X},
	adsurl = {https://ui.adsabs.harvard.edu/abs/2020IJMPA..3550075R}
}

@ARTICLE{Kawamura2006CQGra,
	author = {{Kawamura}, Seiji and {Nakamura}, Takashi and {Ando}, Masaki and {Seto}, Naoki and {Tsubono}, Kimio and {Numata}, Kenji and {Takahashi}, Ryuichi and {Nagano}, Shigeo and {Ishikawa}, Takehiko and {Musha}, Mitsuru and {Ueda}, Ken-ichi and {Sato}, Takashi and {Hosokawa}, Mizuhiko and {Agatsuma}, Kazuhiro and {Akutsu}, Tomotada and {Aoyanagi}, Koh-suke and {Arai}, Koji and {Araya}, Akito and {Asada}, Hideki and {Aso}, Yoichi and {Chiba}, Takeshi and {Ebisuzaki}, Toshikazu and {Eriguchi}, Yoshiharu and {Fujimoto}, Masa-Katsu and {Fukushima}, Mitsuhiro and {Futamase}, Toshifumi and {Ganzu}, Katsuhiko and {Harada}, Tomohiro and {Hashimoto}, Tatsuaki and {Hayama}, Kazuhiro and {Hikida}, Wataru and {Himemoto}, Yoshiaki and {Hirabayashi}, Hisashi and {Hiramatsu}, Takashi and {Ichiki}, Kiyotomo and {Ikegami}, Takeshi and {Inoue}, Kaiki T. and {Ioka}, Kunihito and {Ishidoshiro}, Koji and {Itoh}, Yousuke and {Kamagasako}, Shogo and {Kanda}, Nobuyuki and {Kawashima}, Nobuki and {Kirihara}, Hiroyuki and {Kiuchi}, Kenta and {Kobayashi}, Shiho and {Kohri}, Kazunori and {Kojima}, Yasufumi and {Kokeyama}, Keiko and {Kozai}, Yoshihide and {Kudoh}, Hideaki and {Kunimori}, Hiroo and {Kuroda}, Kazuaki and {Maeda}, Kei-ichi and {Matsuhara}, Hideo and {Mino}, Yasushi and {Miyakawa}, Osamu and {Miyoki}, Shinji and {Mizusawa}, Hiromi and {Morisawa}, Toshiyuki and {Mukohyama}, Shinji and {Naito}, Isao and {Nakagawa}, Noriyasu and {Nakamura}, Kouji and {Nakano}, Hiroyuki and {Nakao}, Kenichi and {Nishizawa}, Atsushi and {Niwa}, Yoshito and {Nozawa}, Choetsu and {Ohashi}, Masatake and {Ohishi}, Naoko and {Ohkawa}, Masashi and {Okutomi}, Akira and {Oohara}, Kenichi and {Sago}, Norichika and {Saijo}, Motoyuki and {Sakagami}, Masaaki and {Sakata}, Shihori and {Sasaki}, Misao and {Sato}, Shuichi and {Shibata}, Masaru and {Shinkai}, Hisaaki and {Somiya}, Kentaro and {Sotani}, Hajime and {Sugiyama}, Naoshi and {Tagoshi}, Hideyuki and {Takahashi}, Tadayuki and {Takahashi}, Hirotaka and {Takahashi}, Ryutaro and {Takano}, Tadashi and {Tanaka}, Takahiro and {Taniguchi}, Keisuke and {Taruya}, Atsushi and {Tashiro}, Hiroyuki and {Tokunari}, Masao and {Tsujikawa}, Shinji and {Tsunesada}, Yoshiki and {Yamamoto}, Kazuhiro and {Yamazaki}, Toshitaka and {Yokoyama}, Jun'ichi and {Yoo}, Chul-Moon and {Yoshida}, Shijun and {Yoshino}, Taizoh},
	title = "{The Japanese space gravitational wave antenna{\textemdash}DECIGO}",
	journal = {Classical and Quantum Gravity},
	year = 2006,
	month = apr,
	volume = {23},
	number = {8},
	pages = {S125-S131},
	doi = {10.1088/0264-9381/23/8/S17},
	adsurl = {https://ui.adsabs.harvard.edu/abs/2006CQGra..23S.125K}
}

@article{Harry_2006,
	doi = {10.1088/0264-9381/23/15/008},
	url = {https://dx.doi.org/10.1088/0264-9381/23/15/008},
	year = {2006},
	month = {jul},
	publisher = {},
	volume = {23},
	number = {15},
	pages = {4887},
	author = {Harry, Gregory M and Fritschel, Peter and Shaddock, Daniel A and Folkner, William and Phinney, E Sterl},
	title = {Laser interferometry for the Big Bang Observer},
	journal = {Classical and Quantum Gravity}
}

@ARTICLE{Cutler2006PhRvD,
	author = {{Cutler}, Curt and {Harms}, Jan},
	title = "{Big Bang Observer and the neutron-star-binary subtraction problem}",
	journal = {\prd},
	year = 2006,
	month = feb,
	volume = {73},
	number = {4},
	eid = {042001},
	pages = {042001},
	doi = {10.1103/PhysRevD.73.042001},
	archivePrefix = {arXiv},
	eprint = {gr-qc/0511092},
	primaryClass = {gr-qc},
	adsurl = {https://ui.adsabs.harvard.edu/abs/2006PhRvD..73d2001C}
}

@ARTICLE{Peters1963,
	author = {{Peters}, P.~C. and {Mathews}, J.},
	title = "{Gravitational Radiation from Point Masses in a Keplerian Orbit}",
	journal = {Physical Review},
	year = 1963,
	month = jul,
	volume = {131},
	number = {1},
	pages = {435-440},
	doi = {10.1103/PhysRev.131.435},
	adsurl = {https://ui.adsabs.harvard.edu/abs/1963PhRv..131..435P}
}

@ARTICLE{Wagg2022ApJS,
	author = {{Wagg}, T. and {Breivik}, K. and {de Mink}, S.~E.},
	title = "{LEGWORK: A Python Package for Computing the Evolution and Detectability of Stellar-origin Gravitational-wave Sources with Space-based Detectors}",
	journal = {\apjs},
	year = 2022,
	month = jun,
	volume = {260},
	number = {2},
	eid = {52},
	pages = {52},
	doi = {10.3847/1538-4365/ac5c52},
	archivePrefix = {arXiv},
	eprint = {2111.08717},
	primaryClass = {astro-ph.HE},
	adsurl = {https://ui.adsabs.harvard.edu/abs/2022ApJS..260...52W}
}

@ARTICLE{Yagi2011PhRvD,
	author = {{Yagi}, Kent and {Seto}, Naoki},
	title = "{Detector configuration of DECIGO/BBO and identification of cosmological neutron-star binaries}",
	journal = {\prd},
	year = 2011,
	month = feb,
	volume = {83},
	number = {4},
	eid = {044011},
	pages = {044011},
	doi = {10.1103/PhysRevD.83.044011},
	archivePrefix = {arXiv},
	eprint = {1101.3940},
	primaryClass = {astro-ph.CO},
	adsurl = {https://ui.adsabs.harvard.edu/abs/2011PhRvD..83d4011Y}
}

@ARTICLE{Yagi2017PhRvD,
	author = {{Yagi}, Kent and {Seto}, Naoki},
	title = "{Erratum: Detector configuration of DECIGO/BBO and identification of cosmological neutron-star binaries [Phys. Rev. D 83, 044011 (2011)]}",
	journal = {\prd},
	year = 2017,
	month = may,
	volume = {95},
	number = {10},
	eid = {109901},
	pages = {109901},
	doi = {10.1103/PhysRevD.95.109901},
	adsurl = {https://ui.adsabs.harvard.edu/abs/2017PhRvD..95j9901Y}
}

@ARTICLE{Sun2024AA,
	author = {{Sun}, Mengfei and {Li}, Jin and {Cao}, Shuo and {Liu}, Xiaolin},
	title = "{Deep learning forecasts of cosmic acceleration parameters from DECi-hertz Interferometer Gravitational-wave Observatory}",
	journal = {\aap},
	year = 2024,
	month = feb,
	volume = {682},
	eid = {A177},
	pages = {A177},
	doi = {10.1051/0004-6361/202347221},
	archivePrefix = {arXiv},
	eprint = {2307.16437},
	primaryClass = {astro-ph.GA},
	adsurl = {https://ui.adsabs.harvard.edu/abs/2024A&A...682A.177S}
}

@ARTICLE{Hamers2021RNAAS,
	author = {{Hamers}, Adrian S.},
	title = "{An Improved Numerical Fit to the Peak Harmonic Gravitational Wave Frequency Emitted by an Eccentric Binary}",
	journal = {Research Notes of the American Astronomical Society},
	year = 2021,
	month = nov,
	volume = {5},
	number = {11},
	eid = {275},
	pages = {275},
	doi = {10.3847/2515-5172/ac3d98},
	archivePrefix = {arXiv},
	eprint = {2111.08033},
	primaryClass = {gr-qc},
	adsurl = {https://ui.adsabs.harvard.edu/abs/2021RNAAS...5..275H}
}

@ARTICLE{Manchester2005AJ,
	author = {{Manchester}, R.~N. and {Hobbs}, G.~B. and {Teoh}, A. and {Hobbs}, M.},
	title = "{The Australia Telescope National Facility Pulsar Catalogue}",
	journal = {\aj},
	year = 2005,
	month = apr,
	volume = {129},
	number = {4},
	pages = {1993-2006},
	doi = {10.1086/428488},
	archivePrefix = {arXiv},
	eprint = {astro-ph/0412641},
	primaryClass = {astro-ph},
	adsurl = {https://ui.adsabs.harvard.edu/abs/2005AJ....129.1993M}
}

@ARTICLE{Kurban2026AA,
	author = {{Kurban}, Abdusattar and {Zhou}, Xia and {Wang}, Na and {Huang}, Yong-Feng and {Yan}, Wenming and {Yuan}, Jianping and {Esamdin}, Ali and {Wang}, Yu-Bin and {Wen}, Zhigang and {Yuen}, Rai},
	title = "{Detectability of continuous gravitational waves from planetary-mass companions orbiting compact stars}",
	journal = {\aap},
	year = 2026,
	month = apr,
	volume = {708},
	eid = {A182},
	pages = {A182},
	doi = {10.1051/0004-6361/202557134},
	archivePrefix = {arXiv},
	eprint = {2604.05425},
	primaryClass = {astro-ph.HE},
	adsurl = {https://ui.adsabs.harvard.edu/abs/2026A&A...708A.182K}
}

@ARTICLE{2011ApJS..194...39F,
	author = {{Fischer}, T. and {Sagert}, I. and {Pagliara}, G. and {Hempel}, M. and {Schaffner-Bielich}, J. and {Rauscher}, T. and {Thielemann}, F.-K. and {K{\"a}ppeli}, R. and {Mart{\'\i}nez-Pinedo}, G. and {Liebend{\"o}rfer}, M.},
	title = "{Core-collapse Supernova Explosions Triggered by a Quark-Hadron Phase Transition During the Early Post-bounce Phase}",
	journal = {\apjs},
	year = 2011,
	month = jun,
	volume = {194},
	number = {2},
	eid = {39},
	pages = {39},
	doi = {10.1088/0067-0049/194/2/39},
	archivePrefix = {arXiv},
	eprint = {1011.3409},
	primaryClass = {astro-ph.HE},
	adsurl = {https://ui.adsabs.harvard.edu/abs/2011ApJS..194...39F}
}

@ARTICLE{2013A&A...558A..50N,
	author = {{Nakazato}, Ken'ichiro and {Sumiyoshi}, Kohsuke and {Yamada}, Shoichi},
	title = "{Stellar core collapse with hadron-quark phase transition}",
	journal = {\aap},
	year = 2013,
	month = oct,
	volume = {558},
	eid = {A50},
	pages = {A50},
	doi = {10.1051/0004-6361/201322231},
	archivePrefix = {arXiv},
	eprint = {1309.3383},
	primaryClass = {astro-ph.HE},
	adsurl = {https://ui.adsabs.harvard.edu/abs/2013A&A...558A..50N}
}

@ARTICLE{2009PhRvL.102h1101S,
	author = {{Sagert}, I. and {Fischer}, T. and {Hempel}, M. and {Pagliara}, G. and {Schaffner-Bielich}, J. and {Mezzacappa}, A. and {Thielemann}, F.-K. and {Liebend{\"o}rfer}, M.},
	title = "{Signals of the QCD Phase Transition in Core-Collapse Supernovae}",
	journal = {\prl},
	year = 2009,
	month = feb,
	volume = {102},
	number = {8},
	eid = {081101},
	pages = {081101},
	doi = {10.1103/PhysRevLett.102.081101},
	archivePrefix = {arXiv},
	eprint = {0809.4225},
	primaryClass = {astro-ph},
	adsurl = {https://ui.adsabs.harvard.edu/abs/2009PhRvL.102h1101S}
}

@ARTICLE{2020PhRvL.125e1102Z,
	author = {{Zha}, Shuai and {O'Connor}, Evan P. and {Chu}, Ming-chung and {Lin}, Lap-Ming and {Couch}, Sean M.},
	title = "{Gravitational-wave Signature of a First-order Quantum Chromodynamics Phase Transition in Core-Collapse Supernovae}",
	journal = {\prl},
	year = 2020,
	month = jul,
	volume = {125},
	number = {5},
	eid = {051102},
	pages = {051102},
	doi = {10.1103/PhysRevLett.125.051102},
	archivePrefix = {arXiv},
	eprint = {2007.04716},
	primaryClass = {astro-ph.HE},
	adsurl = {https://ui.adsabs.harvard.edu/abs/2020PhRvL.125e1102Z}
}

@ARTICLE{2025arXiv251008707Z,
	author = {{Zenati}, Yossef and {Albertus Torres}, Conrado and {Silk}, Joseph and {{\'A}ngeles P{\'e}rez-Garc{\'\i}a}, M.},
	title = "{Neutrino signal from the hadron-quark phase transition in the conversion of Neutron Stars into Quark Stars}",
	journal = {arXiv e-prints},
	year = 2025,
	month = oct,
	eid = {arXiv:2510.08707},
	pages = {arXiv:2510.08707},
	doi = {10.48550/arXiv.2510.08707},
	archivePrefix = {arXiv},
	eprint = {2510.08707},
	primaryClass = {astro-ph.HE},
	adsurl = {https://ui.adsabs.harvard.edu/abs/2025arXiv251008707Z}
}

@ARTICLE{2026PhRvD.113d4002C,
	author = {{Char}, Prasanta and {Biswas}, Bhaskar},
	title = "{Compact object of HESS J1731-347 and its implication on neutron star matter}",
	journal = {\prd},
	year = 2026,
	month = feb,
	volume = {113},
	number = {4},
	eid = {044002},
	pages = {044002},
	doi = {10.1103/f8dq-t7ky},
	archivePrefix = {arXiv},
	eprint = {2408.15220},
	primaryClass = {astro-ph.HE},
	adsurl = {https://ui.adsabs.harvard.edu/abs/2026PhRvD.113d4002C}
}

@ARTICLE{2024ApJ...967..159D,
	author = {{Di Clemente}, Francesco and {Drago}, Alessandro and {Pagliara}, Giuseppe},
	title = "{Is the Compact Object Associated with HESS J1731-347 a Strange Quark Star? A Possible Astrophysical Scenario for Its Formation}",
	journal = {\apj},
	year = 2024,
	month = jun,
	volume = {967},
	number = {2},
	eid = {159},
	pages = {159},
	doi = {10.3847/1538-4357/ad445b},
	archivePrefix = {arXiv},
	eprint = {2211.07485},
	primaryClass = {astro-ph.HE},
	adsurl = {https://ui.adsabs.harvard.edu/abs/2024ApJ...967..159D}
}

@ARTICLE{2025arXiv250802652S,
	author = {{Shirke}, Swarnim and {Maiti}, Rajesh and {Chatterjee}, Debarati},
	title = "{PSR J0614-3329: A NICER case for Strange Quark Stars}",
	journal = {arXiv e-prints},
	year = 2025,
	month = aug,
	eid = {arXiv:2508.02652},
	pages = {arXiv:2508.02652},
	doi = {10.48550/arXiv.2508.02652},
	archivePrefix = {arXiv},
	eprint = {2508.02652},
	primaryClass = {astro-ph.HE},
	adsurl = {https://ui.adsabs.harvard.edu/abs/2025arXiv250802652S}
}

@ARTICLE{2020ApJ...892L...3A,
	author = {{Abbott}, B.~P. and {Abbott}, R. and {Abbott}, T.~D. and {Abraham}, S. and {Acernese}, F. and {Ackley}, K. and {Adams}, C. and {Adhikari}, R.~X. and {Adya}, V.~B. and {Affeldt}, C. and {Agathos}, M. and {Agatsuma}, K. and {Aggarwal}, N. and {Aguiar}, O.~D. and {Aiello}, L. and {Ain}, A. and {Ajith}, P. and {Allen}, G. and {Allocca}, A. and {Aloy}, M.~A. and {Altin}, P.~A. and {Amato}, A. and {Anand}, S. and {Ananyeva}, A. and {Anderson}, S.~B. and {Anderson}, W.~G. and {Angelova}, S.~V. and {Antier}, S. and {Appert}, S. and {Arai}, K. and {Araya}, M.~C. and {Areeda}, J.~S. and {Ar{\`e}ne}, M. and {Arnaud}, N. and {Aronson}, S.~M. and {Arun}, K.~G. and {Ascenzi}, S. and {Ashton}, G. and {Aston}, S.~M. and {Astone}, P. and {Aubin}, F. and {Aufmuth}, P. and {AultONeal}, K. and {Austin}, C. and {Avendano}, V. and {Avila-Alvarez}, A. and {Babak}, S. and {Bacon}, P. and {Badaracco}, F. and {Bader}, M.~K.~M. and {Bae}, S. and {Baird}, J. and {Baker}, P.~T. and {Baldaccini}, F. and {Ballardin}, G. and {Ballmer}, S.~W. and {Bals}, A. and {Banagiri}, S. and {Barayoga}, J.~C. and {Barbieri}, C. and {Barclay}, S.~E. and {Barish}, B.~C. and {Barker}, D. and {Barkett}, K. and {Barnum}, S. and {Barone}, F. and {Barr}, B. and {Barsotti}, L. and {Barsuglia}, M. and {Barta}, D. and {Bartlett}, J. and {Bartos}, I. and {Bassiri}, R. and {Basti}, A. and {Bawaj}, M. and {Bayley}, J.~C. and {Baylor}, A.~C. and {Bazzan}, M. and {B{\'e}csy}, B. and {Bejger}, M. and {Belahcene}, I. and {Bell}, A.~S. and {Beniwal}, D. and {Benjamin}, M.~G. and {Berger}, B.~K. and {Bergmann}, G. and {Bernuzzi}, S. and {Berry}, C.~P.~L. and {Bersanetti}, D. and {Bertolini}, A. and {Betzwieser}, J. and {Bhandare}, R. and {Bidler}, J. and {Biggs}, E. and {Bilenko}, I.~A. and {Bilgili}, S.~A. and {Billingsley}, G. and {Birney}, R. and {Birnholtz}, O. and {Biscans}, S. and {Bischi}, M. and {Biscoveanu}, S. and {Bisht}, A. and {Bitossi}, M. and {Bizouard}, M.~A. and {Blackburn}, J.~K. and {Blackman}, J. and {Blair}, C.~D. and {Blair}, D.~G. and {Blair}, R.~M. and {Bloemen}, S. and {Bobba}, F. and {Bode}, N. and {Boer}, M. and {Boetzel}, Y. and {Bogaert}, G. and {Bondu}, F. and {Bonnand}, R. and {Booker}, P. and {Boom}, B.~A. and {Bork}, R. and {Boschi}, V. and {Bose}, S. and {Bossilkov}, V. and {Bosveld}, J. and {Bouffanais}, Y. and {Bozzi}, A. and {Bradaschia}, C. and {Brady}, P.~R. and {Bramley}, A. and {Branchesi}, M. and {Brau}, J.~E. and {Breschi}, M. and {Briant}, T. and {Briggs}, J.~H. and {Brighenti}, F. and {Brillet}, A. and {Brinkmann}, M. and {Brockill}, P. and {Brooks}, A.~F. and {Brooks}, J. and {Brown}, D.~D. and {Brunett}, S. and {Buikema}, A. and {Bulik}, T. and {Bulten}, H.~J. and {Buonanno}, A. and {Buskulic}, D. and {Buy}, C. and {Byer}, R.~L. and {Cabero}, M. and {Cadonati}, L. and {Cagnoli}, G. and {Cahillane}, C. and {Calder{\'o}n Bustillo}, J. and {Callister}, T.~A. and {Calloni}, E. and {Camp}, J.~B. and {Campbell}, W.~A. and {Canepa}, M. and {Cannon}, K.~C. and {Cao}, H. and {Cao}, J. and {Carapella}, G. and {Carbognani}, F. and {Caride}, S. and {Carney}, M.~F. and {Carullo}, G. and {Casanueva Diaz}, J. and {Casentini}, C. and {Caudill}, S. and {Cavagli{\`a}}, M. and {Cavalier}, F. and {Cavalieri}, R. and {Cella}, G. and {Cerd{\'a}-Dur{\'a}n}, P. and {Cesarini}, E. and {Chaibi}, O. and {Chakravarti}, K. and {Chamberlin}, S.~J. and {Chan}, M. and {Chao}, S. and {Charlton}, P. and {Chase}, E.~A. and {Chassande-Mottin}, E. and {Chatterjee}, D. and {Chaturvedi}, M. and {Chatziioannou}, K. and {Cheeseboro}, B.~D. and {Chen}, H.~Y. and {Chen}, X. and {Chen}, Y. and {Cheng}, H.-P. and {Cheong}, C.~K. and {Chia}, H.~Y. and {Chiadini}, F. and {Chincarini}, A. and {Chiummo}, A. and {Cho}, G. and {Cho}, H.~S.},
	title = "{GW190425: Observation of a Compact Binary Coalescence with Total Mass {\ensuremath{\sim}} 3.4 M$_{☉}$}",
	journal = {\apjl},
	year = 2020,
	month = mar,
	volume = {892},
	number = {1},
	eid = {L3},
	pages = {L3},
	doi = {10.3847/2041-8213/ab75f5},
	archivePrefix = {arXiv},
	eprint = {2001.01761},
	primaryClass = {astro-ph.HE},
	adsurl = {https://ui.adsabs.harvard.edu/abs/2020ApJ...892L...3A}
}

@ARTICLE{2021PhRvL.126p2702B,
	author = {{Bombaci}, I. and {Drago}, A. and {Logoteta}, D. and {Pagliara}, G. and {Vida{\~n}a}, I.},
	title = "{Was GW190814 a Black Hole-Strange Quark Star System?}",
	journal = {\prl},
	year = 2021,
	month = apr,
	volume = {126},
	number = {16},
	eid = {162702},
	pages = {162702},
	doi = {10.1103/PhysRevLett.126.162702},
	archivePrefix = {arXiv},
	eprint = {2010.01509},
	primaryClass = {nucl-th},
	adsurl = {https://ui.adsabs.harvard.edu/abs/2021PhRvL.126p2702B}
}

@ARTICLE{2021PhRvC.103b5808D,
	author = {{Dexheimer}, V. and {Gomes}, R.~O. and {Kl{\"a}hn}, T. and {Han}, S. and {Salinas}, M.},
	title = "{GW190814 as a massive rapidly rotating neutron star with exotic degrees of freedom}",
	journal = {\prc},
	year = 2021,
	month = feb,
	volume = {103},
	number = {2},
	eid = {025808},
	pages = {025808},
	doi = {10.1103/PhysRevC.103.025808},
	archivePrefix = {arXiv},
	eprint = {2007.08493},
	primaryClass = {astro-ph.HE},
	adsurl = {https://ui.adsabs.harvard.edu/abs/2021PhRvC.103b5808D}
}

@ARTICLE{2021ApJ...910...62Z,
	author = {{Zhou}, Xia and {Li}, Ang and {Li}, Bao-An},
	title = "{R-mode Stability of GW190814's Secondary Component as a Supermassive and Superfast Pulsar}",
	journal = {\apj},
	year = 2021,
	month = mar,
	volume = {910},
	number = {1},
	eid = {62},
	pages = {62},
	doi = {10.3847/1538-4357/abe538},
	archivePrefix = {arXiv},
	eprint = {2011.11934},
	primaryClass = {astro-ph.HE},
	adsurl = {https://ui.adsabs.harvard.edu/abs/2021ApJ...910...62Z}
}

@ARTICLE{2025RAA....25e5016Y,
	author = {{Yuan}, Ya-Jing and {Zhou}, Xia},
	title = "{Thermal Evolution of the Central Compact Object in HESS J1731{\ensuremath{-}}347 as Evidence for a Color-flavor-locked Strange Star}",
	journal = {Research in Astronomy and Astrophysics},
	year = 2025,
	month = may,
	volume = {25},
	number = {5},
	eid = {055016},
	pages = {055016},
	doi = {10.1088/1674-4527/adce4e},
	adsurl = {https://ui.adsabs.harvard.edu/abs/2025RAA....25e5016Y}
}

@ARTICLE{1994PhRvL..73.1328C,
	author = {{Cottingham}, W.~N. and {Kalafatis}, D. and {Vinh Mau}, R.},
	title = "{Brown dwarfs, quark stars, and quark-hadron phase transition}",
	journal = {\prl},
	year = 1994,
	month = sep,
	volume = {73},
	number = {10},
	pages = {1328-1331},
	doi = {10.1103/PhysRevLett.73.1328},
	adsurl = {https://ui.adsabs.harvard.edu/abs/1994PhRvL..73.1328C}
}

@ARTICLE{1987PhLB..192...71O,
	author = {{Olinto}, Angela V.},
	title = "{On the conversion of neutron stars into strange stars}",
	journal = {Physics Letters B},
	year = 1987,
	month = jun,
	volume = {192},
	number = {1-2},
	pages = {71-75},
	doi = {10.1016/0370-2693(87)91144-0},
	adsurl = {https://ui.adsabs.harvard.edu/abs/1987PhLB..192...71O}
}

@ARTICLE{1998ChPhL..15..934X,
	author = {{Xu}, Ren-xin and {Qiao}, Guo-jun},
	title = "{`Bare' Strange Stars Might Not Be Bare}",
	journal = {Chinese Physics Letters},
	year = 1998,
	month = dec,
	volume = {15},
	number = {12},
	pages = {934-936},
	doi = {10.1088/0256-307X/15/12/026},
	archivePrefix = {arXiv},
	eprint = {astro-ph/9811197},
	primaryClass = {astro-ph},
	adsurl = {https://ui.adsabs.harvard.edu/abs/1998ChPhL..15..934X}
}

@ARTICLE{2006APh....25..212X,
	author = {{Xu}, R.~X.},
	title = "{To probe into pulsar{\textquoteright}s interior through gravitational waves}",
	journal = {Astroparticle Physics},
	year = 2006,
	month = apr,
	volume = {25},
	number = {3},
	pages = {212-219},
	doi = {10.1016/j.astropartphys.2006.01.004},
	archivePrefix = {arXiv},
	eprint = {astro-ph/0511612},
	primaryClass = {astro-ph},
	adsurl = {https://ui.adsabs.harvard.edu/abs/2006APh....25..212X}
}

@ARTICLE{2012RAA....12..813H,
	author = {{Horvath}, J.~E.},
	title = "{The nature of the companion of PSR J1719-1438: a white dwarf or an exotic object?}",
	journal = {Research in Astronomy and Astrophysics},
	year = 2012,
	month = jul,
	volume = {12},
	number = {7},
	pages = {813-816},
	doi = {10.1088/1674-4527/12/7/009},
	archivePrefix = {arXiv},
	eprint = {1205.1410},
	primaryClass = {astro-ph.HE},
	adsurl = {https://ui.adsabs.harvard.edu/abs/2012RAA....12..813H}
}

@ARTICLE{1995ApJ...440..815D,
	author = {{Dai}, Zigao and {Peng}, Qiuhe and {Lu}, Tan},
	title = "{The Conversion of Two-Flavor to Three-Flavor Quark Matter in a Supernova Core}",
	journal = {\apj},
	year = 1995,
	month = feb,
	volume = {440},
	pages = {815},
	doi = {10.1086/175316},
	adsurl = {https://ui.adsabs.harvard.edu/abs/1995ApJ...440..815D}
}

@ARTICLE{1996PhRvL..77.1210C,
	author = {{Cheng}, K.~S. and {Dai}, Z.~G.},
	title = "{Conversion of Neutron Stars to Strange Stars as a Possible Origin of {\ensuremath{\gamma}}-Ray Bursts}",
	journal = {\prl},
	year = 1996,
	month = aug,
	volume = {77},
	number = {7},
	pages = {1210-1213},
	doi = {10.1103/PhysRevLett.77.1210},
	archivePrefix = {arXiv},
	eprint = {astro-ph/9510073},
	primaryClass = {astro-ph},
	adsurl = {https://ui.adsabs.harvard.edu/abs/1996PhRvL..77.1210C}
}

@ARTICLE{2009PhRvL.103a1101B,
	author = {{Bauswein}, A. and {Janka}, H.-T. and {Oechslin}, R. and {Pagliara}, G. and {Sagert}, I. and {Schaffner-Bielich}, J. and {Hohle}, M.~M. and {Neuh{\"a}user}, R.},
	title = "{Mass Ejection by Strange Star Mergers and Observational Implications}",
	journal = {\prl},
	year = 2009,
	month = jul,
	volume = {103},
	number = {1},
	eid = {011101},
	pages = {011101},
	doi = {10.1103/PhysRevLett.103.011101},
	archivePrefix = {arXiv},
	eprint = {0812.4248},
	primaryClass = {astro-ph},
	adsurl = {https://ui.adsabs.harvard.edu/abs/2009PhRvL.103a1101B}
}

\end{document}